\documentclass[%
 aip,
 pof,%
 amsmath,amssymb,
reprint,%
]{revtex4-1}

\usepackage{graphicx}
\usepackage{dcolumn}
\usepackage{bm}
\usepackage[mathlines]{lineno}
\usepackage{color}
\usepackage{hyperref}

\usepackage{makecell}
\begin{document}
\preprint{AIP/123-QED}

\title{Numerical study of cavitation bubble dynamics in a flowing tube}
\author{Nian Wang}
\affiliation{State Key Laboratory of Engines, Tianjin University, Tianjin, 300350, China.}%
\author{Odumuyiwa A. Odumosu}
\affiliation{State Key Laboratory of Engines, Tianjin University, Tianjin, 300350, China.}%
\author{Tianyou Wang}
\affiliation{State Key Laboratory of Engines, Tianjin University, Tianjin, 300350, China.}%
\affiliation{National Industry-Education Platform of Energy Storage, Tianjin University, Tianjin, 300350, China}
\author{Zhizhao Che}
\email{chezhizhao@tju.edu.cn}
 \affiliation{State Key Laboratory of Engines, Tianjin University, Tianjin, 300350, China.}
 \affiliation{National Industry-Education Platform of Energy Storage, Tianjin University, Tianjin, 300350, China}
\date{\today}

\begin{abstract}
Cavitation in tubes is a common occurrence in nature and engineering applications. Previous studies of cavitation bubble dynamics mainly consider bubbles in stagnant-water tubes, but the dynamics of cavitation bubbles in tubes with flow is not clear. This study investigates the dynamics of cavitation bubbles in tubes with flow by numerical simulations. The results show that, unlike bubbles in stagnant-water tubes, bubbles under the combined effects of water inflow and tube wall confinement exhibit asymmetric behavior along the axis of the tube. The inflow suppresses the development of the bubble interface near the tube inlet, causing that side of the interface to move with the inflow. In contrast, the expansion and contraction of the bubble and the generation of liquid jets occur on the side near the outlet. This feature results in significant asymmetry in the bubble interface, therefore we introduce a skewness parameter to characterize the difference in length between the left and right parts of the bubble during the bubble evolution. The evolution of the bubble significantly affects the mass flow rate at the outlet of the tube, and even leads to backflow during the bubble contraction process.
\end{abstract}


\maketitle
\onecolumngrid
\section{Introduction}\label{sec:1}
Cavitation is the process in which a liquid undergoes a phase transition into vapor due to a local pressure drop below the saturation vapor pressure. As the cavitation process is extremely transient, cavitation bubbles generate a very high energy density at the moment of collapse, resulting in significant effects on the surrounding environment, examples include cavitation erosion \cite{Bilus2020, Du2021}, noise \cite{LiLINMIN2023, Si2023}, and luminescence \cite{Nanzai2023, Yusof2022}. The occurrence of cavitation is prevalent in engineering applications such as fluid machinery \cite{Ebrahimi2021, JiaX2023, Wang2022Num}, materials engineering \cite{Nikolaev2018, Rezk2021, Xu2013}, biomedicine \cite{Brennen2015, Chen2011, Hosny2013}, and surface cleaning \cite{Chahine2016, LiPan2022Bubble, Zhong2022}.

The dynamic behaviors of a single cavitation bubble in a free domain can be described by the Rayleigh equation \cite{Rayleigh1917}. The effects of liquid viscosity, surface tension, and non-condensable gas on bubble dynamics can be incorporated into the Rayleigh equation to derive the Rayleigh-Plesset equation \cite{Plesset1949}. Fluid compressibility has a significant effect on bubble evolution as well \cite{Keller1980, Zhang2023A}. In a free domain, a single cavitation bubble typically maintains a spherical shape during its evolution. However, in practical applications, cavitation bubbles often form at various boundaries, such as rigid walls \cite{Wang2022, Xu2023}, elastic boundaries \cite{Brujan2001, Reese2023}, free surfaces \cite{LiT2019Interaction, Zhang2018}, and mixed boundaries \cite{CuiP2016, LiS2023Vertical, LiS2019Jet, ZhangAM2015}. The presence of such boundaries often leads to bubble deformation and the generation of liquid jets \cite{Cui2020, CuiR2023, Gonzalez2020, Zhang2021}.

In practical applications, cavitation often occurs within flowing liquids, and the presence of flow makes bubble dynamics more complex. In flowing liquids, bubbles experience a reduction in length along the flow direction, causing them to transition from a spherical to an ellipsoidal shape. Additionally, significant asymmetry in the shape of the bubble appears on both sides along the flow direction \cite{Zhou2022}. Flow directed perpendicular to a rigid wall significantly influences the behavior of bubbles near the wall. With decreasing the initial distance from the bubble to the wall, the bubble exhibits different dynamic characteristics, including pinching, formation of a needle-like high-speed jet, and departure from the wall during the collapse phase \cite{Mnich2024}. Bubbles in shear flow near a rigid wall are subject to the combined effects of wall-induced lift and shear-induced lift forces, resulting in a helical trajectory. With increasing the initial distance from the bubble to the wall, the migration distance of the bubble in the vertical direction first increases and then decreases \cite{Su2024}.

Tubes are common boundaries where cavitation phenomena occur. In a stagnant-water tube, bubbles are constrained by the tube wall, leading to a significant increase in their lifetime \cite{Aghdam2015}. As the bubble contracts, two counter jets emerge from the axial ends toward the midpoint of the tube. When the counter jets meet, a ring jet is formed, dividing the bubble into two \cite{Ni2012}. The size of the tube and the distance of the bubble from the midpoint of the tube can significantly affect the bubble behaviors \cite{Wang2024WallConf}. When the tube diameter is large enough, the difference between the oscillation period of the bubble and that of the bubble in free domains is small \cite{LiJie2024Collapsing}. As the tube deviates from the midpoint, the evolution of bubbles exhibits asymmetry. Axial deviation from the midpoint results in a pumping effect, where liquid near one side of the bubble flows toward the other side \cite{Ory2000, Yuan1999}. Radial deviation from the midpoint causes the bubble to migrate toward the nearby tube wall when its equivalent radius is smaller than the tube radius, or toward the far tube wall when its equivalent radius is larger than the tube radius \cite{Wang2019Exp}.When bubbles are in profiled tubes, more complex dynamics characteristics such as asymmetric jets \cite{ren22} and bubble splitting \cite{nagargoje23} occur.

In tubes with flows, the behavior of bubbles becomes more complex due to the combined effects of the incoming flow and the tube wall. In a vertically oriented tube with initial flow, under the combined action of the incoming flow at the tube inlet and buoyancy, a bubble forms an upward jet at its bottom and pierces through the bubble to form a ring \cite{Luo2022}. When bubbles inside the tube are relatively large, the flow of liquid around the bubbles causes the middle part of the bubble to form a cylindrical shape with a constant radius. The movement speed of the bubble is faster than the flow speed of the liquid inside the tube \cite{Feng2009}. When bubbles inside the tube are small, unlike large bubbles in tubes, the behavior of small bubbles is significantly influenced by the relative sizes of the bubbles and the tube. When the bubble is very small (with a bubble equivalent radius less than one-fourth of the tube radius), its movement speed is always equal to the flow speed of the liquid at the axis of the tube \cite{Feng2010}.

Although several studies have been conducted some on cavitation bubbles in tubes with flow, the understanding of this process remains incomplete. Particularly, the mechanisms of the combined effects of the incoming flow and the tube wall on cavitation bubbles still need to be elucidated, such as the asymmetric development of bubbles on both sides along the tube axis. Therefore, this study numerically simulates the dynamics of cavitation bubbles in tubes with flow, focusing on the asymmetric shape changes of bubbles during their evolution. Additionally, we investigate the effects of tube diameter, length, incoming flow velocity, and liquid viscosity on bubble evolution.

\section{Numerical methods}\label{sec:2}
\subsection{Governing equations}\label{sec:2.1}
The simulation is based on OpenFOAM \cite{Weller1998} with the Finite Volume Method (FVM) to solve the compressible Navier-Stokes equations and the Volume-of-Fluid (VOF) method to capture the gas-liquid interface. The simulation uses a pressure-based compressible two-phase flow solver, solving equations including the continuity equation, momentum equation, energy equation, and volume fraction equation:
\begin{equation}\label{eq:01}
\frac{\partial \rho }{\partial t}+\nabla \cdot (\rho \mathbf{u})=0
\end{equation}
\begin{equation}\label{eq:02}
\frac{\partial \rho \mathbf{u}}{\partial t}+\nabla \cdot (\rho \mathbf{uu})=-\nabla p+\nabla \cdot \mathbf{\tau} +{{\mathbf{f}}_{\sigma }}
\end{equation}
\begin{equation}\label{eq:03}
\frac{\partial \rho T}{\partial t}+\nabla \cdot (\rho \mathbf{u}T)+\frac{1}{{{c}_{v}}}\left[ \frac{\partial \left( \rho k \right)}{\partial t}+\nabla \cdot (\rho \mathbf{u}k)+\nabla \cdot (p\mathbf{u}) \right]=\nabla \cdot ({{\alpha }_\text{eff}}\nabla T)
\end{equation}
\begin{equation}\label{eq:04}
\frac{\partial {{\alpha }_{l}}}{\partial t}+\nabla \cdot \left( {{\alpha }_{l}}\mathbf{u} \right)+\nabla \cdot \left( {{\alpha }_{l}}{{\alpha }_{v}}{{\mathbf{u}}_{r}} \right)={{\alpha }_{l}}{{\alpha }_{v}}\left( \frac{1}{{{\rho }_{v}}}\frac{d{{\rho }_{v}}}{dt}-\frac{1}{{{\rho }_{l}}}\frac{d{{\rho }_{l}}}{dt} \right)+{{\alpha }_{l}}\nabla \cdot \mathbf{u}
\end{equation}
where $\rho$ is the fluid density, $\mathbf{u}$ is the velocity vector, $p$ is the pressure, $\tau =\mu [ \nabla \mathbf{u} +\nabla {{\mathbf{u}}^{T}}-2/3(\nabla \cdot \mathbf{u})\mathbf{I} ]$ is the viscous stress tensor, $\mu$ is the dynamic viscosity, $\mathbf{I}$ is the unit tensor, $\mathbf{f}_\sigma$ represents the source term resulting from surface tension, calculated using the continuum surface force model \cite{Brackbill1992}, $T$ is the temperature, $c_v$ is the specific heat capacity, $\alpha_\text{eff}$ is thermal diffusivity, $k=0.5{{\left| \mathbf{u} \right|}^{2}}$ is the specific kinetic energy, $\alpha$ is the phase volume fraction, and ${{\mathbf{u}}_{{r}}}$ is the relative velocity between the two phases. The subscripts $l$ and $v$ represent the liquid and gas phases, respectively.

The physical parameters at the gas-liquid interface, such as density $\rho$ and dynamic viscosity $\mu$, are also calculated from volume fractions.
\begin{equation}\label{eq:05}
\beta ={{\alpha }_{l}}{{\beta }_{l}}+{{\alpha }_{v}}{{\beta }_{v}}
\end{equation}
where $\beta_l$ and $\beta_v$ represent the physical parameters of the liquid phase and the gas phase, respectively.

The liquid follows the Tait equation of state \cite{Koch2016}
\begin{equation}\label{eq:06}
	{{\rho }_{l}}={{\rho }_{\text{ref}}}{{\left( \frac{p+B}{{{p}_{\text{ref}}}+B} \right)}^{\frac{1}{{{n}_{l}}}}}
\end{equation}
where $\rho_\text{ref}$ and $p_\text{ref}$ are the reference density and pressure, respectively, and $\rho_\text{ref} = 998$ kg/m$^3$, $p_\text{ref} = 101325$ Pa, $n_l = 7.15$, and $B = 3.046\times {{10}^{8}}$ Pa, respectively.

The gas follows the ideal gas equation of state
\begin{equation}\label{eq:07}
	{{\rho }_{v}}=\frac{p}{RT}
\end{equation}
\subsection{Numerical implementation}\label{sec:2.2}
\begin{figure}
  \centering
  \includegraphics[scale=0.9]{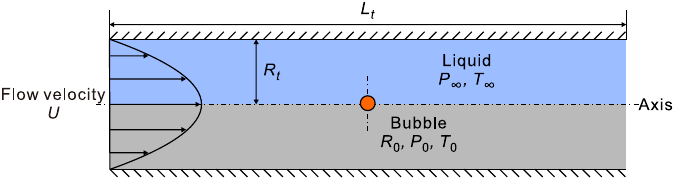}
  \caption{Schematic diagram of the numerical model for a cavitation bubble in a flowing tube.}\label{fig:01}
\end{figure}

The simulation settings of this study are shown in Figure \ref{fig:01}. An axisymmetric setting was used to reduce the computational cost. The length and the radius of the tube are $L_t$ and $R_t$, respectively. An adiabatic no-slip boundary condition was used at the tube wall. A parabolic velocity inlet boundary condition was used at the left boundary of the tube
\begin{equation}\label{eq:08}
	U\left( r \right)={{U}_{m}}\left[ 1-{{\left( \frac{r}{{{R}_{t}}} \right)}^{2}} \right]
\end{equation}
where $U_m$ is the maximum flow velocity at the axis of the tube. A pressure outlet boundary condition was used at the right boundary of the tube.

In this study, the bubble was positioned at the midpoint of the tube initially, and it was set to be spherical with a radius, pressure, and temperature of $R_0$, $P_0$, and $T_0$, respectively. The liquid inside the tube was initially set to $P_\infty$ and $T_\infty$ for pressure and temperature, respectively. At the initial moment of the simulation, the velocity profile of the liquid flow throughout the tube was set to be parabolic, being the same as the inlet flow distribution.

To analyze the combined effects of inflow and tube wall on the evolution of bubbles, we defined several dimensionless parameters that affect bubble dynamics. The dimensionless tube diameter and length are
\begin{equation}\label{eq:09}
	{{R}_{t}}^{*}=\frac{{{R}_{t}}}{{{R}_{m}}}
\end{equation}
\begin{equation}\label{eq:10}
	{{L}_{t}}^{*}=\frac{{{L}_{t}}}{{{R}_{m}}}
\end{equation}
The dimensionless inflow velocity is
\begin{equation}\label{eq:11}
	{{U}^{*}}=\frac{{\bar{U}}}{{{v}_{m}}}
\end{equation}
where $\bar{U}$ is the average flow velocity at the tube inlet
\begin{equation}\label{eq:12}
	\bar{U}=\frac{{{U}_{m}}}{2}
\end{equation}
The dimensionless time is
\begin{equation}\label{eq:13}
	{{t}^{*}}=\frac{t}{{{t}_{m}}}
\end{equation}
where
\begin{equation}\label{eq:14}
	{{t}_{m}}={{R}_{m}}\sqrt{\frac{\rho_{l} }{{{P}_{\infty }}}}
\end{equation}
where $R_m$ and $v_m$ denote the theoretical maximum radius and maximum velocity of the bubble, respectively, which were derived from the solution of the Rayleigh equation \cite{Rayleigh1917}
\begin{equation}\label{eq:15}
	1-\frac{R_{m}^{3}}{R_{0}^{3}}+\frac{{{P}_{0}}}{{{P}_{\infty }}}\ln \left( \frac{R_{m}^{3}}{R_{0}^{3}} \right)=0
\end{equation}
\begin{equation}\label{eq:16}
	{{v}_{m}}^{2}=\frac{2{{P}_{0}}}{3{{\rho }_{l}}}{{e}^{{({{P}_{\infty }}-{{P}_{0}})}/{{{P}_{0}}}}}-\frac{2{{P}_{\infty }}}{3{{\rho }_{l}}}
\end{equation}
Here, $t_m$ is on the same order of magnitude as the bubble period, and can reasonably represent the expansion and contraction process of the bubble.

\subsection{Experimental validation}\label{sec:2.3}
\begin{figure}
  \centering
  \includegraphics[scale=0.9]{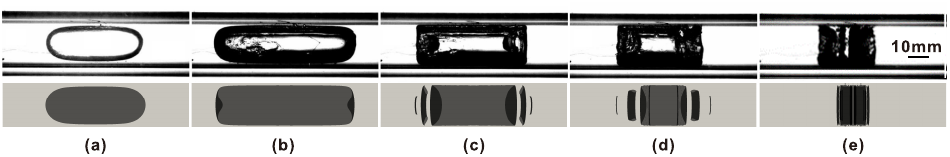}
  \caption{Numerical model validation results (the top row is the experimental results in Ref.~\citenum{Wang2019Exp}, and the bottom row is the numerical simulation results, respectively). The dimensionless times in the figures are: (a) $\hat{t} = -0.3$, (b) $\hat{t} = 0$, (c) $\hat{t} = 0.7$, (d) $\hat{t} = 0.8$, and (e) $\hat{t} = 1$. Here $\hat{t}=\left( t-{{t}_{\max }} \right)/\left( {{t}_{coll}}-{{t}_{\max }} \right)$, where $t_\text{max}$ is the time of maximum bubble size, and $t_\text{coll}$ is the bubble collapse time. The experimental images are reprinted with permission from Wang et al., Int. J. Multiphase Flow 121, 103096 (2019). Copyright 2019 Elsevier.}
  \label{fig:02}
\end{figure}

To verify the accuracy of the numerical model of bubble dynamics in the tube, we compared the numerical simulation results with experimental results. The parameters of the simulation are consistent with the cavitation bubble experiment described in Ref.~\citenum{Wang2019Exp}. To simulate the influence of the water tank on the bubble in the experiment, cylindrical liquid pools with a diameter and height of $8R_t$ were placed at both ends of the tube. More details of the simulations can be found in Ref.~\citenum{Wang2024WallConf}. As shown in Figure \ref{fig:02}, in comparison to experimental results, the numerical model effectively captures the evolution of the bubble and reproduces the primary dynamic features of the bubble.

\subsection{Grid independence study}\label{sec:2.4}
To assess result accuracy, a study on grid independence was carried out for the numerical model. In this study, the bubble was positioned at the midpoint of the tube and assumed to be spherical, with $R_0 = 5$ mm, $P_0 = 1.3\times10^6$ Pa, and $T_0 = 400$ K. The pressure and temperature of the liquid in the tube are $P_\infty = 101325$ Pa, and $T_\infty = 297$ K, respectively. The average inlet velocity is $\bar{U} = 1.5$ m/s, and the sizes of the tube are $R_t = 15$ mm and $L_t = 160$ mm. The theoretical parameters of the bubble are $R_m = 18.7$ mm, $v_m = 17$ m/s, and $t_m = 1.9$ ms, which means $R_t^* = 0.802$, $L_t^* = 8.556$, and $U^* = 0.088$, respectively.

\begin{figure}
  \centering
  \includegraphics[scale=0.9]{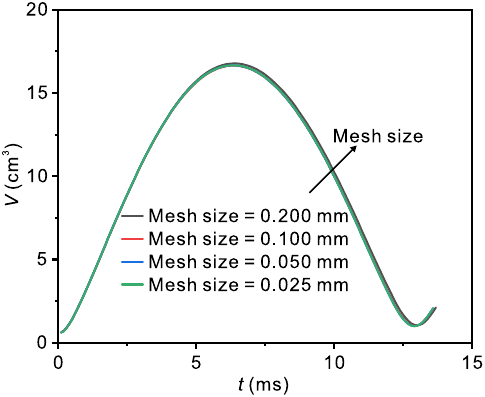}
  \caption{Grid independence study for the bubble size using varying mesh sizes. Here, $R_t^* = 0.802$, $L_t^* = 8.556$, and $U^* = 0.088$.}\label{fig:03}
\end{figure}

We performed grid independence validation with four different grid resolutions. The variation of bubble volume $V$ with time $t$ for mesh sizes of 0.2, 0.1, 0.05, and 0.025 mm is shown in Figure \ref{fig:03}. The results indicate that refining the mesh size from 0.05 to 0.025 mm has a negligible effect on bubble evolution. Consequently, a mesh size of 0.05 mm was adopted for numerical simulations in this study.

\section{Results and discussion}\label{sec:3}
\subsection{Bubble evolution process}\label{sec:3.1}
\begin{figure}
  \centering
  \includegraphics[scale=0.9]{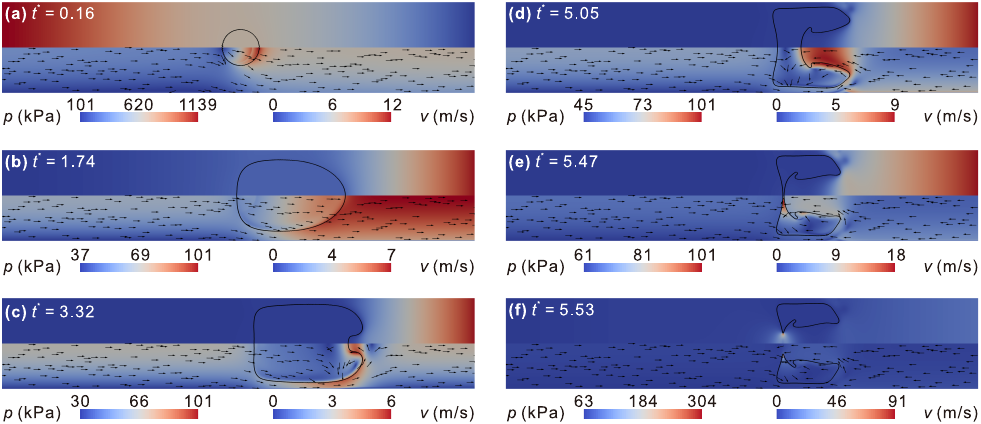}
  \caption{Bubble shape, pressure distribution (upper parts of the figures), and velocity field (lower parts of the figures) during bubble evolution. Here $R_t^* = 0.802$, $L_t^* = 8.556$, and $U^* = 0.088$. (Multimedia view).}\label{fig:04}
\end{figure}

The evolution process of a bubble in a tube with flow under a typical condition ($R_t^*$ = 0.802, $L_t^*$ = 8.556, and $U^*$ = 0.088) is shown in Figure \ref{fig:04} (Multimedia view). Initially, driven by the pressure difference between the interior and exterior of the bubble, the bubble expands rapidly. However, the inflow at the tube inlet restricts the expansion of the left interface of the bubble, leading to a significantly faster expansion velocity on the right side of the bubble, as shown in Figure \ref{fig:04}(a). As the bubble enlarges, the influence of the tube wall on constraining the bubble expands stronger. The bubble expands much more slowly along the radial direction of the tube compared to the right side, causing the right side of the bubble to become more elongated. The expansion of the bubble pushes the liquid inside the tube to flow out from the tube outlet, leading to an increased flow velocity at the tube outlet. Due to the bubble occupying most of the cross-sectional area of the tube, the incoming flow can only flow through the gap between the bubble and the tube wall toward the outlet, causing an increase in velocity at the gap. As the bubble further expands, the pressure inside the bubble gradually becomes lower than that at the tube outlet, as shown in Figure \ref{fig:04}(b). When the bubble reaches its maximum size, driven by the pressure difference between the tube outlet and the inside of the bubble, the right side of the bubble contracts along the axis of the tube, forming a liquid jet directed toward the midpoint of the tube, as shown in Figure \ref{fig:04}(c). With the development of the liquid jet, its velocity and diameter gradually increase, while the left side of the bubble gradually caves in under the influence of the incoming flow. The incoming flow passes through the gap between the bubble and the tube wall, then, driven by the right-hand-side liquid jet, it deflects and flows toward the left, as shown in Figure \ref{fig:04}(d)-(e). Eventually, the incoming flow meets the right-hand-side jet, piercing the bubble and forming a ring, as shown in Figure \ref{fig:04}(f).

\begin{figure}
  \centering
  \includegraphics[scale=0.8]{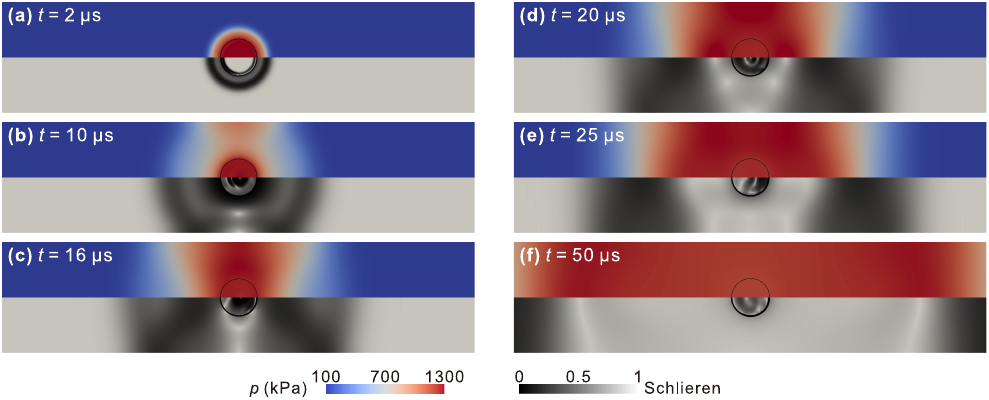}
  \caption{Pressure wave propagation in the early expansion stage (upper parts of the figures) and its numerical schlieren images (lower parts of the figures). Here $R_t^* = 0.802$, $L_t^* = 8.556$, and $U^* = 0.088$.}\label{fig:05}
\end{figure}

The propagation of pressure wave within the tube during the initial expansion of the bubble and its numerical schlieren are shown in Figure \ref{fig:05}. The rapid expansion of the bubble generates a spherical pressure wave that propagates outward, as shown in Figure \ref{fig:05}(a). Subsequently, the pressure wave is reflected by the tube wall, forming a high-pressure region between the bubble and the tube wall, as shown in Figures \ref{fig:05}(b)-(c). The pressure wave continues to propagate along the axial direction of the tube, forming high-pressure areas on both sides of the bubble, as shown in Figures \ref{fig:05}(d)-(f). Due to the inlet flow velocity (1.5 m/s) being much lower than the propagation speed of the pressure wave in water (1500 m/s), the inflow has no significant effect on the propagation of the pressure wave in the water.

\begin{figure}
  \centering
  \includegraphics[scale=0.9]{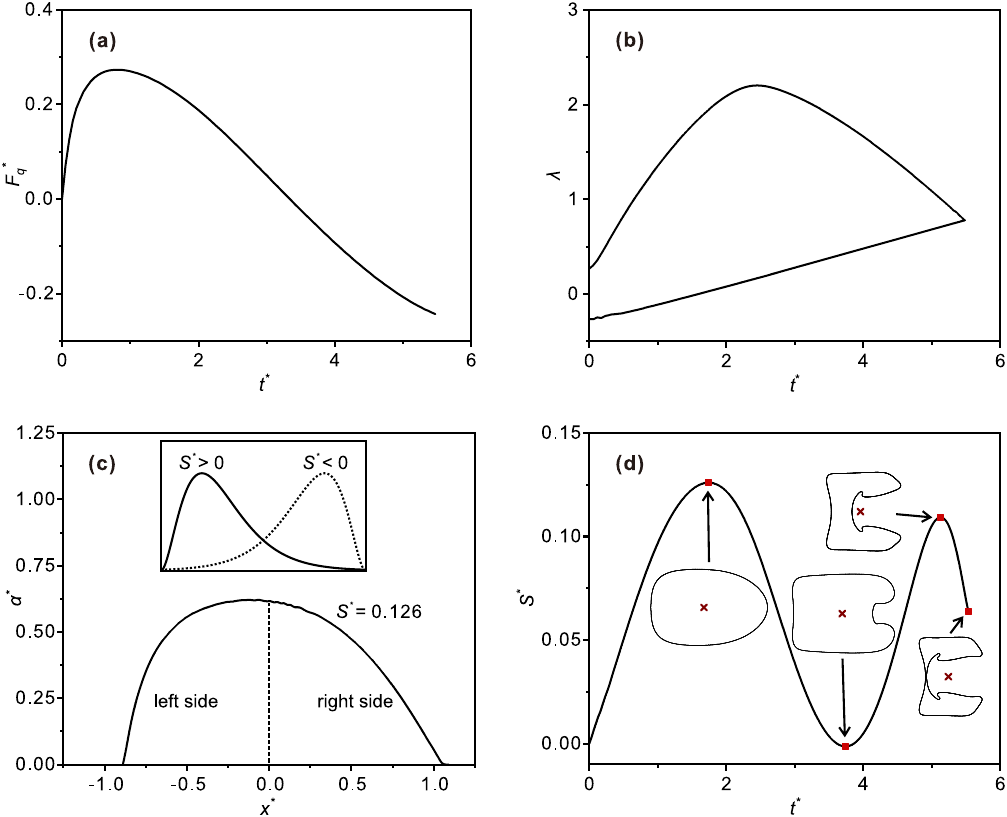}
  \caption{(a) Variation of mass flow rate ratio $F_q^*$; (b) Variation of interface positions on both sides of the bubble $\lambda$; (c) Mass distribution of the left and right parts of the bubble (divided by the bubble centroid) at a typical moment ($t^* = 1.74$), here ${{x}^{*}}=\left( x-\bar{x} \right)/{{R}_{m}}$ and ${{\alpha }^{*}}={{A}_{b}}/{{A}_{t}}$, where $x$ is the axial coordinate of the bubble distribution, $\bar{x}$ is the axial coordinate of the bubble's center of mass, $A_b$ is the cross-sectional area of the bubble in the corresponding axial coordinate and $A_t$ is the cross-sectional area of the tube, respectively; (d) Variation of bubble skewness $S^*$, the insets show the bubble shape and center of mass position. Here $R_t^* = 0.802$, $L_t^* = 8.556$, and $U^* = 0.088$.}\label{fig:06}
\end{figure}

Figure \ref{fig:04} illustrates that the evolution of the bubble significantly influences the flow at the outlet of the tube: the bubble expansion drives the liquid to exit at the tube outlet, resulting in an increased outlet flow rate (as shown in Figure \ref{fig:04}(b)); the bubble contraction results in a decreased outlet flow rate, and even reverses the flow direction, causing backflow (as shown in Figure \ref{fig:04}(d)). To characterize the influence of bubble evolution on the flow at the tube outlet, we define a dimensionless mass flow rate ratio
\begin{equation}\label{eq:17}
	{{F}_{q}}^{*}=\frac{{{{\dot{m}}}_{out}}-{{{\dot{m}}}_{\text{in}}}}{{{\rho}_{l}}\frac{4\pi {{R}_{m}}^{3}}{3{{t}_{m}}}}
\end{equation}
where ${{\dot{m}}_\text{out}}$ is the instantaneous mass flow rate at the tube outlet (positive for outflow, negative for inflow), and ${{\dot{m}}_\text{in}}$ is the mass flow rate at the tube inlet. As shown in Figure \ref{fig:06}(a), during the process from the initiation of bubble expansion to the meeting of the bubble's two side interfaces, $F_q^*$ first increases and then decreases to negative. The reason for this is the expansion of the bubble causes the expulsion of liquid from the tube outlet, resulting in an increase in the mass flow rate at the tube outlet. As the bubble expansion rate decreases, the mass flow rate at the tube outlet gradually decreases. When the bubble contracts, the outlet mass flow rate is less than the inlet mass flow rate and backflow occurs at the tube outlet. This results in $F_q^* < 0$, and the magnitude of $F_q^*$ increases as the right-hand-side liquid jet develops.

During the evolution of the bubble, the inflow at the tube inlet induces asymmetry on both sides of the bubble. To characterize the degree of asymmetry on both sides of the bubble, the dimensionless interface position is defined
\begin{equation}\label{eq:18}
	\lambda =\frac{{{x}_\text{{int}}}}{{{R}_{m}}}
\end{equation}
where $x_\text{int}$ is the instantaneous axial coordinate of the interface positions on both sides of the bubble.As shown in Figure \ref{fig:06}(b), there is a significant asymmetry in the positions of the interfaces on both sides of the bubble. The upper part of the curve is the right-hand-side interface of the bubble, while the lower part is the left-hand-side interface. The intersection point of the curve is the meeting position of the interfaces on both sides of the bubble. The left-hand-side interface of the bubble is suppressed by the incoming flow and undergoes minimal expansion, gradually moving toward the outlet of the tube. Conversely, the right-hand-side interface of the bubble moves rapidly toward the tube outlet driven by the pressure difference. Its movement speed gradually decreases, and after reaching the farthest position, the right-hand-side interface begins to contract. The contraction speed gradually increases until it meets the left-hand-side interface. Because of the pushing effect of the incoming flow, the meeting point of the bubble's two side interfaces is located to the right of the bubble's initial position.

The shape of the bubble in Figure \ref{fig:04} indicates that the left-hand side of the bubble undergoes minimal shape change under the suppression of the inflow, while the right-hand side undergoes significant deformation during expansion and contraction. Since the bubble's expansion and contraction mainly occur in the axial direction, the axial length of the bubble undergoes significant variation during this process. The mass distribution of the two parts of the bubble (divided by the bubble's center of mass) at a typical moment ($t^* = 1.74$, with the bubble shape shown in Figure \ref{fig:04}(b)) is shown in Figure \ref{fig:06}(c). Compared to the left half, the distribution of the bubble mass in the right half is longer, and the proportion of the bubble occupying the cross-section of the tube is smaller, indicating that the bubble in the right half is more elongated. This indicates that during the evolution of the bubble, the left and right halves, which have the same volume, are different in length. To characterize the difference in length between the left and right sides of the bubble, we define the dimensionless skewness parameter
\begin{equation}\label{eq:19}
	{{S}^{*}}=\frac{{{\mu }_{3}}}{{{\sigma }^{3}}}
\end{equation}
where $\mu _3$ and $\sigma$ are the third central moment and standard deviation of the bubble's distribution along the axial direction, respectively, which are calculated as
\begin{equation}\label{eq:20}
	{{\mu }_{3}}=\frac{\int{{{\left( x-\bar{x} \right)}^{3}}\left( 1-{{\alpha }_{l}} \right)}dV}{\int{\left( 1-{{\alpha }_{l}} \right)}dV}
\end{equation}
\begin{equation}\label{eq:21}
	\sigma ={{\left[ \frac{\int{{{\left( x-\bar{x} \right)}^{2}}\left( 1-{{\alpha }_{l}} \right)}dV}{\int{\left( 1-{{\alpha }_{l}} \right)}dV} \right]}^{1/2}}
\end{equation}
Hence, $S^*$ represents the degree of asymmetry in the lengths of the two parts of the bubble. $S^* > 0$ indicates that the axial distance from the bubble's center of mass to its right-hand side is longer, meaning the right half of the bubble is more elongated (as shown by the solid line in the inset of Figure \ref{fig:06}(c)). $S^* < 0$ indicates that the axial distance from the bubble's center of mass to its left-hand side is longer, indicating that the left half of the bubble is more elongated (as shown by the dashed line in the inset of Figure \ref{fig:06}(c)). An increase in the magnitude of $S^*$ indicates an increase in the difference in the lengths of the bubble's two parts.

The variation of the bubble skewness $S^*$ with time is shown in Figure \ref{fig:06}(d). During the process from initial expansion to the meeting of the bubble's two side interfaces, the skewness exhibits two oscillations. In the initial stage, the bubble primarily expands to the right, causing the shape of the right half of the bubble to become more elongated. The difference in length between the two halves increases as the bubble expands. Under the restriction of the pressure difference between the outlet of the tube and inside the bubble, the expansion speed of the bubble's right-hand-side interface gradually decreases to negative. During this process, the right-hand-side interface of the bubble gradually flattens and even starts to concave, resulting in a reduction in the length difference between the two parts of the bubble. As the liquid jet develops, its diameter gradually increases, causing the annular bubble on the right-hand side to become thinner, thereby increasing the length difference between the two parts of the bubble. Before the bubble's two side interfaces meet, the development of the jet causes the left half of the bubble to become annular, reducing the length difference between the two halves of the bubble.

\subsection{Effect of tube diameter}\label{sec:3.2}
\begin{figure}
  \centering
  \includegraphics[scale=0.9]{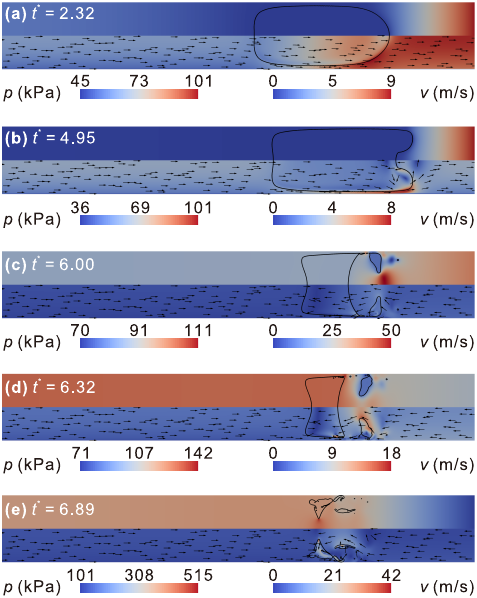}
  \caption{Bubble shape, pressure distribution, and velocity field during bubble evolution with a small diameter. Here, $R_t^* = 0.588$, $L_t^* = 8.556$, and $U^* = 0.088$. (Multimedia view).}\label{fig:07}
\end{figure}

To investigate the impact of tube diameter on bubble evolution, Figure \ref{fig:07} (Multimedia view) illustrates the bubble evolution process under a small tube diameter ($R_t^* = 0.588$). Compared to a large tube diameter ($R_t^* = 0.802$, as shown in Figure \ref{fig:04}), the bubble is more elongated during the expansion process due to stronger restriction from the tube wall (as shown in Figure \ref{fig:07}(a)). At the time of the maximum volume, the bubble deviates more from a spherical shape (compared to Figure \ref{fig:04}(c), as shown in Figure \ref{fig:07}(b)). In the process of bubble contraction, the influence of the right liquid jet and the viscous resistance of the tube wall leads to bubble breakup near the wall on the right-hand side (as shown in Figure \ref{fig:07}(c)). Because the bubble almost completely occupies the cross-section of the tube during contraction, limiting the flow of liquid on both sides of the bubble through the gap between the bubble and the tube wall, the inflow and liquid jet compress the bubble to a very small volume before the bubble's two side interfaces meet, leading to increased pressure inside the bubble that exceeds the tube outlet pressure (as shown in Figure \ref{fig:07}(d)). At the moment when the bubble's two side interfaces meet, its volume decreases significantly, and the degree of deformation increases noticeably (compared to Figure \ref{fig:04}(f), as shown in Figure \ref{fig:07}(e)).

\begin{figure}
  \centering
  \includegraphics[scale=0.9]{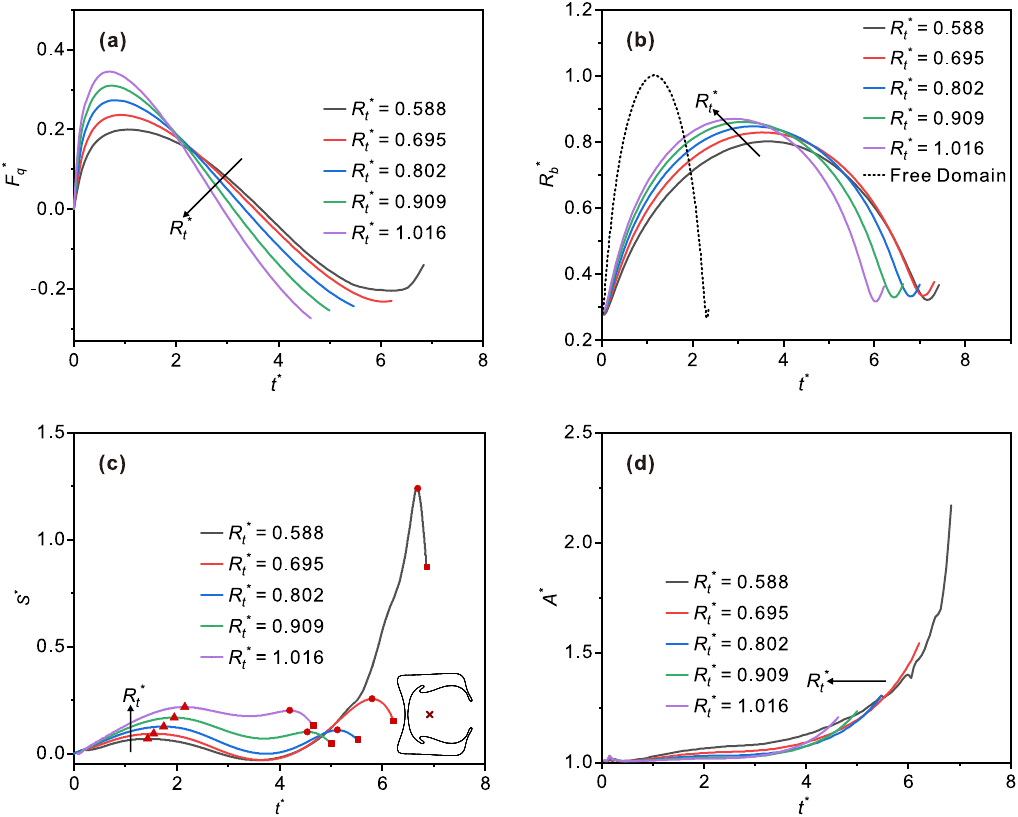}
  \caption{Effect of the tube diameter $R_t^*$ on the bubble shape with $L_t^* = 8.556$ and $U^* = 0.088$: (a) Variation of mass flow rate ratio $F_q^*$; (b) Variation of bubble size $R_b^*$; (c) Variation of bubble skewness $S^*$; (d) Variation of bubble deformation $A^*$.}\label{fig:08}
\end{figure}

To analyze the impact of $R_t^*$ on the mass flow rate at the outlet, the variation of the mass flow rate ratio $F_q^*$ with time for different $R_t^*$ is shown in Figure \ref{fig:08}(a). As $R_t^*$ increases, the maximum value of $F_q^*$ increases, and when the bubble's two side interfaces meet, the magnitude of $F_q^*$ increases (note that at this time $F_q^* < 0$). This is because as the tube diameter increases, the restriction of the tube wall to the bubble weakens and the rates of bubble expansion and contraction increase. When $R_t^*$ = 0.588, the pressure inside the bubble exceeds the pressure at the tube outlet before the two side interfaces meet, resulting in slower backflow at the tube outlet (as shown in Figure \ref{fig:07}(d)).

The mass flow rate at the tube outlet is directly influenced by the variation of the bubble size. To describe the change in bubble size in the evolution process, a dimensionless bubble size parameter is defined
\begin{equation}\label{eq:22}
	{{R}_{b}}^{*}=\frac{\sqrt[3]{3V/4\pi }}{{{R}_{m}}}
\end{equation}
where $V$ is the instantaneous volume of the bubble. The impact of $R_t^*$ on the bubble size $R_b^*$ is shown in Figure \ref{fig:08}(b). The dashed line in the Figure 8(b) shows the variation of bubble size in free domains with time, which is obtained by solving the Rayleigh equation \cite{Rayleigh1917}. By comparing with the variation of bubble size in free domains, we can see that the restriction imposed by the tube and the inlet flow leads to a reduction in the maximum size of the bubble and an extension of the bubble period. As $R_t^*$ increases, the maximum size of the bubble increases, while its lifetime decreases. This is because the fluid flow inside the tube is restricted by the tube wall, which resists the evolution of the bubble, resulting in a reduction of the maximum size of the bubble and a prolongation of the bubble period. The restriction exerted by the tube wall on the bubble diminishes with the increase in $R_t^*$, leading to an increased maximum volume during bubble expansion and a reduction of the bubble evolution process. This is consistent with the results in Ref.~\citenum{LiJie2024Collapsing}. When the tube diameter is small ($R_t^*$ = 0.588), during the contraction process, the inlet flow and liquid jet compress the bubble to a very small volume, resulting in a decrease in the minimum volume of the bubble (as shown in Figure \ref{fig:07}(e)). Due to the increased restriction from the tube wall, the bubble becomes more elongated, causing the bubble to be closer to the tube outlet, reducing the amount of liquid between the bubble and the tube outlet, thereby accelerating the bubble's contraction process.

Under the constraint of the incoming flow at the tube inlet, the deformation on the left-hand side of the bubble is minimal, and the expansion and contraction of the bubble mainly occur on the right-hand side, resulting in a significant asymmetry in the shape of the bubble on both sides. To study the effect of $R_t^*$ on the asymmetry of the bubble shape, the change of the bubble skewness $S^*$ with time under different $R_t^*$ is shown in Figure \ref{fig:08}(c). As $R_t^*$ increases, the peak value of the skewness in the expansion process (the first peak, as shown by the triangle symbol in Figure \ref{fig:08}(c)) increases. The peak value in the contraction process (the second peak, as shown by the circular symbol in Figure \ref{fig:08}(c)) decreases first and then increases. The skewness $S^*$ at the moment when the bubble interfaces on both sides meet (as shown by the square symbol in Figure \ref{fig:08}(c)) decreases first and then increases. This is because, during the evolution of the bubble, the constraint of the tube wall causes the bubble to primarily expand and contract along the axial direction. When the tube diameter is small, the constraint of the tube wall causes the overall shape of the bubble to be elongated, and the expansion and contraction of the right interface have a minor effect on the difference in lengths between the left and right parts of the bubble. As the tube diameter increases, the weakening of the constraint of the tube wall causes the length of the entire bubble to decrease, and the effect of the expansion and contraction of the right interface on the difference in the lengths of the left and right parts of the bubble is enhanced. However, when the tube diameter is small ($R_t^* < 0.695$), the right-hand side of the bubble forms a slender tip ($R_t^* = 0.695$, as shown in the inset in Figure \ref{fig:08}(c)) and even experiences fragmentation ($R_t^* = 0.588$, as shown in Figures \ref{fig:07}(c)-(e)), which results in an increase in the length of the right part of the bubble, a significant difference in the lengths between the left and right parts of the bubble, and a marked increase in the bubble skewness.

The asymmetrical behavior of the bubble's left and right parts leads to the bubble shape deviating significantly from a spherical shape. To characterize the degree of deformation of the bubble during the evolution process, a dimensionless deformation parameter is defined
\begin{equation}\label{eq:23}
	{{A}^{*}}=\frac{\sqrt{A/4\pi }}{\sqrt[3]{3V/4\pi }}
\end{equation}
where $A$ is the instantaneous surface area of the bubble. $A^*$ characterizes the extent of deviation of the bubble's shape from a perfect sphere. If the bubble is a perfect sphere, $A^*$ = 1, while larger $A^*$ values indicate greater deviation from spherical. The effect of the tube diameter $R_t^*$ on the bubble deformation $A^*$ is shown in Figure \ref{fig:08}(d). During the process from the initiation of bubble expansion to the meeting of the bubble's two side interfaces, $A^*$ gradually increases, indicating an increasing degree of deformation during the bubble's evolution. As $R_t^*$ increases, $A^*$ decreases gradually, indicating that the restriction of the bubble by the tube wall weakens with increasing tube diameter, and the bubble's shape gradually approaches a sphere. When the tube diameter is small ($R_t^* = 0.588$), the deformation $A^*$ does not increase monotonically with time; instead, it exhibits significant fluctuations (around $t^* = 6.05$ in Figure \ref{fig:08}(d)), which is because bubble fragmentation during the contraction process, leading to a sudden decrease in $A^*$.

\begin{figure}
  \centering
  \includegraphics[scale=0.9]{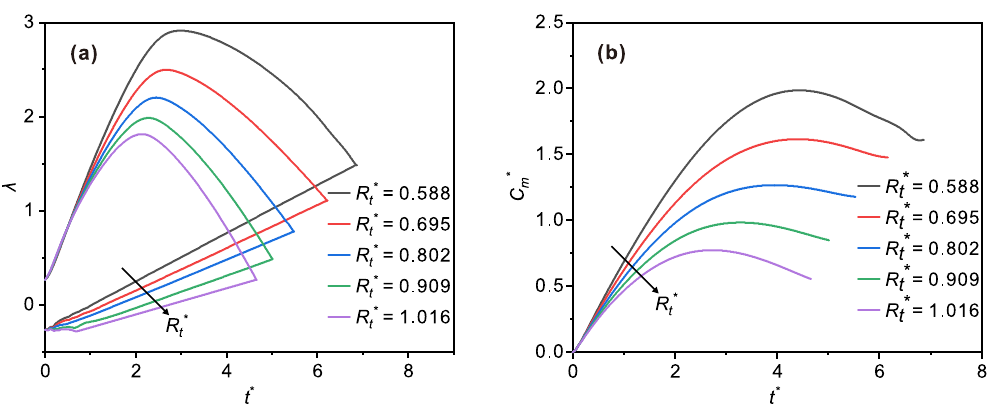}
  \caption{Effect of the tube diameter $R_t^*$ on the bubble position with $L_t^* = 8.556$ and $U^* = 0.088$: (a) Variation of interface positions on the two sides of the bubble $\lambda$; (b) Variation of bubble's center of mass position $C_m^*$. }\label{fig:09}
\end{figure}

Under the influence of the incoming flow, the positions of the left and right interfaces of the bubble exhibit significant asymmetry. The influence of the tube diameter $R_t^*$ on the positions of the bubble's left and right interfaces $\lambda$ is shown in Figure \ref{fig:09}(a). As the tube diameter $R_t^*$ increases, the speed of the bubble's left interface decreases, while the speed of the right interface decreases during the expansion process and increases during the contraction process. This is because with increasing $R_t^*$, the proportion of the bubble in the tube cross-section decreases, reducing the pushing effect of the incoming flow on the bubble and causing a decrease in the speed of the left interface. And as $R_t^*$ increases, the restraining effect of the tube wall on the bubble decreases, leading to increased radial expansion of the bubble during expansion, thus reducing the axial expansion speed of the right interface. The weakening of the restraining effect of the tube wall also causes an increase in the contraction velocity of the bubble's right interface. The change in the velocity of the bubble's interfaces causes the meeting position of the two interfaces to shift toward the inlet of the tube.

Under the propulsion of the incoming flow, the bubble gradually migrates toward the outlet of the tube. Additionally, the asymmetrical behavior of the bubble's interfaces also influences its axial migration process. To characterize the axial migration of the bubble in its evolution process, a dimensionless bubble's center of mass position parameter is defined
\begin{equation}\label{eq:24}
	{C}_{m}^{*}=\frac{{\bar{x}}}{{{R}_{m}}}
\end{equation}
During the process from the initiation of bubble expansion to the meeting of the bubble's two side interfaces, the variation of $C_m^*$ with time for different $R_t^*$ is shown in Figure \ref{fig:09}(b). During the expansion process, the bubble's center of mass moves from the midpoint of the tube toward the outlet, while during the contraction phase, it moves toward the midpoint of the tube. As $R_t^*$ increases, the final migration distance of the bubble's center of mass decreases. This is because the decreased proportion of the bubble within the tube cross-section due to the increase in $R_t^*$, resulting in a decrease in the driving force exerted by the incoming flow and subsequently reducing the migration distance of the bubble.

\subsection{Effect of tube length}\label{sec:3.3}
\begin{figure}
  \centering
  \includegraphics[scale=0.9]{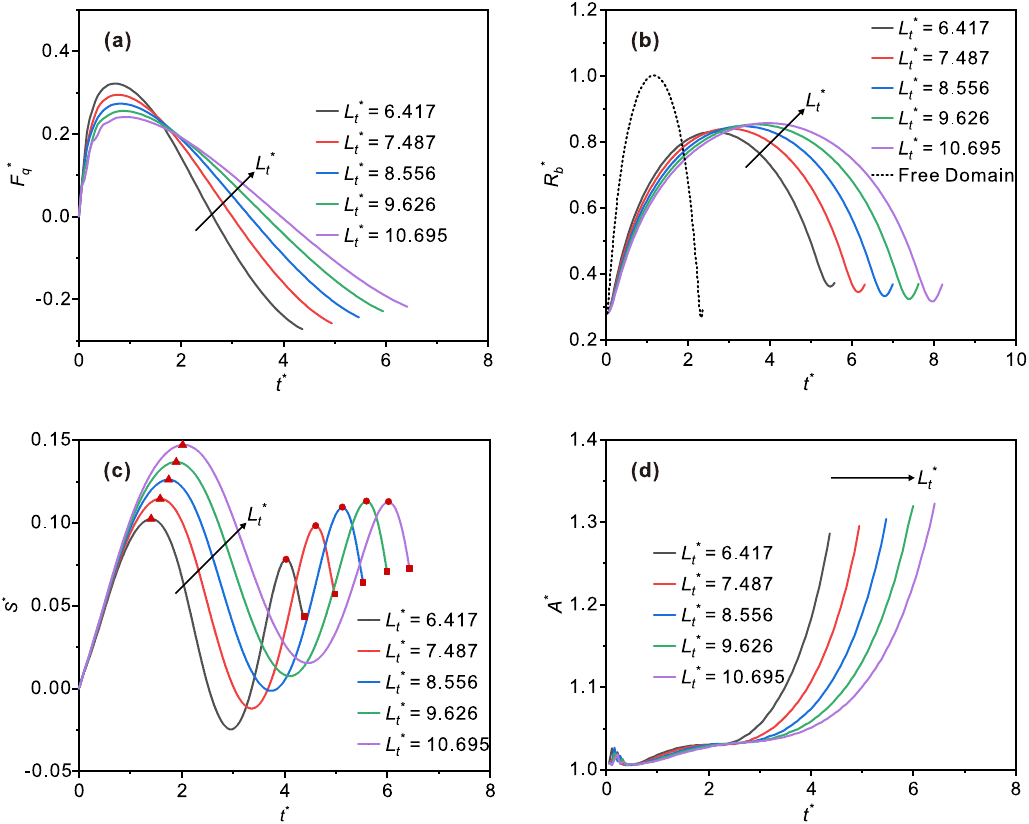}
  \caption{Effect of the tube length $L_t^*$ on the bubble shape with $R_t^* = 0.802$ and $U^* = 0.088$: (a) Variation of mass flow rate ratio $F_q^*$; (b) Variation of bubble size $R_b^*$; (c) Variation of bubble skewness $S^*$; (d) Variation of bubble deformation $A^*$.}\label{fig:10}
\end{figure}

The effect of the tube length $L_t^*$ on the mass flow rate ratio $F_q^*$ is shown in Figure \ref{fig:10}(a). As $L_t^*$ increases, the maximum value of $F_q^*$ gradually decreases, and the magnitude of $F_q^*$ at the moment when the two sides of the bubble meet decreases gradually (note that at this time $F_q^* < 0$). This is because with the increase in tube length, the liquid volume within the tube increases, and the inertia of the liquid between the tube outlet and the bubble increases, which limits the expansion and contraction of the bubble, reducing the rate of expansion and contraction of the bubble and decreasing the degree to which the evolution of the bubble affects the outlet flow rate. Figure \ref{fig:10}(b) further shows the impact of $L_t^*$ on the bubble size $R_b^*$. As $L_t^*$ increases, the bubble maximum size gradually increases, and the bubble lifetime gradually extends. This is because the pressure difference driving the expansion is much greater than the pressure difference driving the contraction. The increase in liquid inertia to some extent suppresses the bubble's contraction, resulting in the maximum size of the bubble increases.

The effect of the tube length $L_t^*$ on the bubble skewness $S^*$ is shown in Figure \ref{fig:10}(c). As $L_t^*$ increases, the peak values of $S^*$ during the expansion process (the first peak, as shown by the triangle symbol in Figure \ref{fig:10}(c)), the contraction process (the second peak, as shown by the circular symbol in Figure \ref{fig:10}(c)), and at the moment when the two sides of the bubble interface meet (as shown by the square symbol in Figure \ref{fig:10}(c)) all increase. This is because the increase in tube length prolongs the expansion process on the bubble's right side. The expansion distance of the right half of the bubble increases, resulting in the length of the right half of the bubble increasing, and the difference in length between the bubble's left and right sides increases. During the process of the bubble contracting to meet at the two sides, the elongation of the tube length prolongs the bubble contraction process. The right jet develops fully, increasing its diameter. The shape of the bubble's right part becomes more elongated, leading to an increase in the difference in length between the two sides of the bubble. The change in the bubble deformation $A^*$ from Figure \ref{fig:10}(d) indicates that as $L_t^*$ increases, the extent of bubble deformation gradually increases. This is because the increase in the tube length prolongs the bubble's evolution process. Under the pushing effect of the incoming flow, the left-hand side of the bubble gradually concaves, while the right-hand side becomes increasingly complex in shape because of the continuous development of the liquid jet.

\begin{figure}
  \centering
  \includegraphics[scale=0.9]{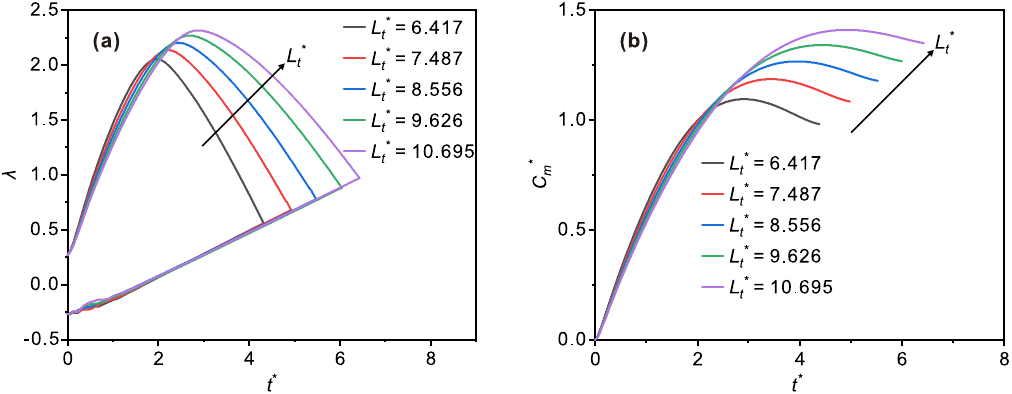}
  \caption{Effect of the tube length $L_t^*$ on the bubble position with $R_t^* = 0.802$ and $U^* = 0.088$: (a) Variation of interface positions on the two sides of the bubble $\lambda$; (b) Variation of bubble's center of mass position $C_m^*$. }\label{fig:11}
\end{figure}

The effect of the tube length $L_t^*$ on the bubble's two side interface positions $\lambda$ is shown in Figure \ref{fig:11}(a). With an increase in $L_t^*$, the expansion and contraction velocities of the bubble's right-hand-side interface gradually decrease due to the increase in the amount of liquid inside the tube. However, the decrease in contraction velocity is more pronounced. Meanwhile, the velocity of movement for the left-hand-side interface remains nearly constant. Consequently, the positions where the two interfaces meet gradually shift toward the outlet of the tube. The effect of the tube length $L_t^*$ on $C_m^*$ is shown in Figure \ref{fig:11}(b). As $L_t^*$ increases, the final migration distance of the bubble's center of mass increases. This is because the increase in tube length prolongs the evolution time of the bubble. Consequently, the pushing action of the incoming flow on the bubble lasts longer, leading to the bubble's final migration distance increase.

\subsection{Effect of inflow velocity}\label{sec:3.4}
\begin{figure}
  \centering
  \includegraphics[scale=0.9]{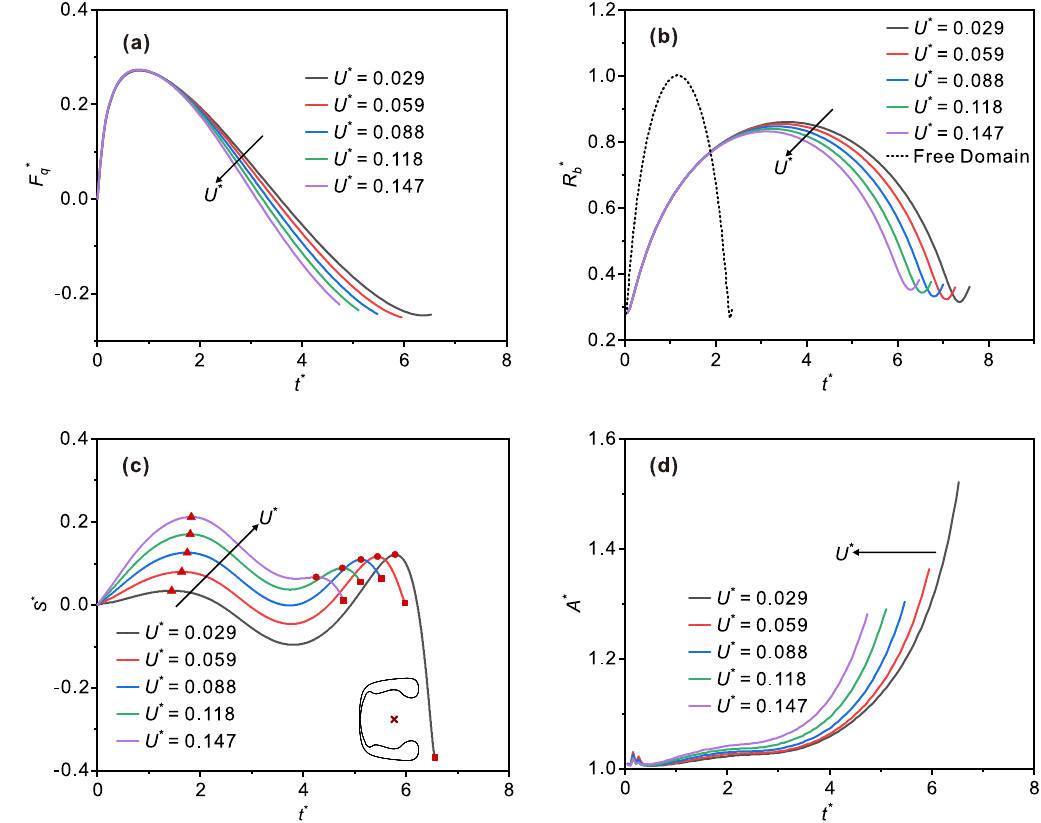}
  \caption{Effect of the inflow velocity $U^*$ on the bubble shape with $R_t^* = 0.802$ and $L_t^* = 8.556$: (a) Variation of mass flow rate ratio $F_q^*$; (b) Variation of bubble size $R_b^*$; (c) Variation of bubble skewness $S^*$; (d) Variation of bubble deformation $A^*$.}\label{fig:12}
\end{figure}

The effect of the inflow velocity $U^*$ on the mass flow rate ratio $F_q^*$ is shown in Figure \ref{fig:12}(a). As $U^*$ increases, the maximum value of $F_q^*$ remains almost constant, and the magnitude of $F_q^*$ at the moment when the bubble's two side interfaces meet decreases (note that at this moment $F_q^* < 0$). This is because as the inflow velocity increases, the indentation of the bubble's left interface under the push of the inflow becomes more pronounced, which leads to the meeting of the two interfaces occurring earlier and results in the under-development of the right-hand-side liquid jet, thereby reducing the outlet mass flow rate. The effect of the inflow velocity $U^*$ on the bubble size $R_b^*$ is shown in Figure \ref{fig:12}(b). As $U^*$ increases, the maximum size of the bubble gradually decreases, and the bubble lifetime shortens. This is because the increase in inflow velocity leads to the bubble being pushed closer to the tube outlet, reducing the amount of liquid between the bubble and the tube outlet. This accelerates the bubble's expansion and contraction processes, with a greater acceleration effect on the contraction process. Consequently, both the maximum size and lifetime of the bubble decrease.

The effect of the inflow velocity $U^*$ on the bubble skewness $S^*$ is shown in Figure \ref{fig:12}(c). With the increase in $U^*$, the peak value of $S^*$ during the expansion (the first peak, as shown by the triangle symbol in Figure \ref{fig:12}(c)) increases, while the peak value during the contraction (the second peak, as shown by the circular symbol in Figure \ref{fig:12}(c)) decreases. The value at the moment when the two sides of the bubble interface meet (as shown by the square symbol in Figure \ref{fig:12}(c)) first decreases then increases, and then decreases again (note that when $U^* = 0.029$, $S^* < 0$). This is because, during the expansion process of the bubble, the increase in flow velocity leads to a flatter motion of the left interface of the bubble, resulting in a decrease in the length of the left half of the bubble and thus an increase in the difference between the lengths of the two parts of the bubble. During the contraction process of the bubble, the increase in flow velocity causes the left interface of the bubble to concave further under the influence of the flow, leading to an increase in the length of the left half of the bubble. Additionally, due to the shortened contraction process of the bubble, the right jet has not fully developed, resulting in a smaller diameter of the right jet and thus a thicker ring of the bubble in the right half. Consequently, the length of the right half of the bubble decreases. These factors collectively reduce the difference in lengths between the left and right parts of the bubble. When the bubble interfaces on both sides meet, a higher inflow velocity leads to increased concavity of the left bubble interface and a decrease in the diameter of the right jet. This causes an increase in the length of the left half of the bubble and a decrease in the length of the right half, reducing the difference in length between the two halves. However, at lower inflow velocities, the left bubble interface does not concave, resulting in the meeting point of the two interfaces being at the far left of the bubble. Additionally, the right jet develops fully, causing an increase in the length of the left half of the bubble. The length of the left half is even greater than that of the right half at $U^* = 0.029$, leading to $S^* < 0$ (as shown in the inset of Figure \ref{fig:12}(c)). The impact of the bubble's asymmetric behavior on its deformation degree under various $U^*$ is shown in Figure \ref{fig:12}(d). As $U^*$ increases, the bubble's deformation gradually decreases. This is because the increase in inflow velocity causes a shorter evolution process of the bubble. The insufficient development of the liquid jet on the right-hand side of the bubble results in a smaller jet diameter, leading to less deformation of the bubble.

\begin{figure}
  \centering
  \includegraphics[scale=0.9]{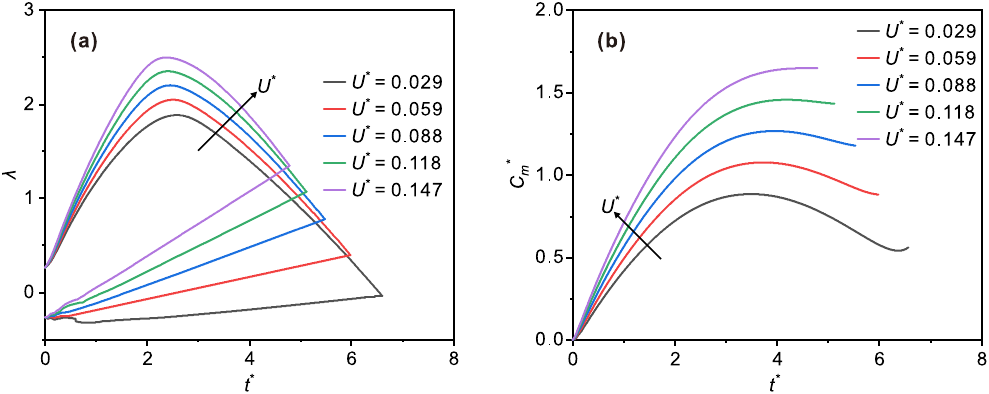}
  \caption{Effect of the inflow velocity $U^*$ on the bubble position with $R_t^* = 0.802$ and $L_t^* = 8.556$: (a) Variation of interface positions on the two sides of the bubble $\lambda$; (b) Variation of bubble's center of mass position $C_m^*$. }\label{fig:13}
\end{figure}

The impact of the inflow velocity $U^*$ on the bubble's interface position on the two sides $\lambda$ is shown in Figure \ref{fig:13}(a). As $U^*$ increases, the movement speed of the bubble's left side interface under the influence of the inflow increases. Meanwhile, the expansion and contraction speeds of the bubble's right-hand side interface both increase. This is because the bubble is closer to the outlet of the tube, the decrease in the mass of fluid on the right-hand side leads to an acceleration of both the expansion and contraction processes of the bubble. The change in the movement speed of the bubble's interfaces causes the meeting position of the two sides to shift toward the outlet of the tube. The variation of the bubble's center of mass position $C_m^*$ with time under different inflow velocities $U^*$ is shown in Figure \ref{fig:13}(b), where the increase in the inflow velocity leads to a greater distance for the bubble's center of mass to eventually migrate under the influence of the inflow.

\subsection{Effect of liquid viscosity}\label{sec:3.5}
\begin{figure}
  \centering
  \includegraphics[scale=0.9]{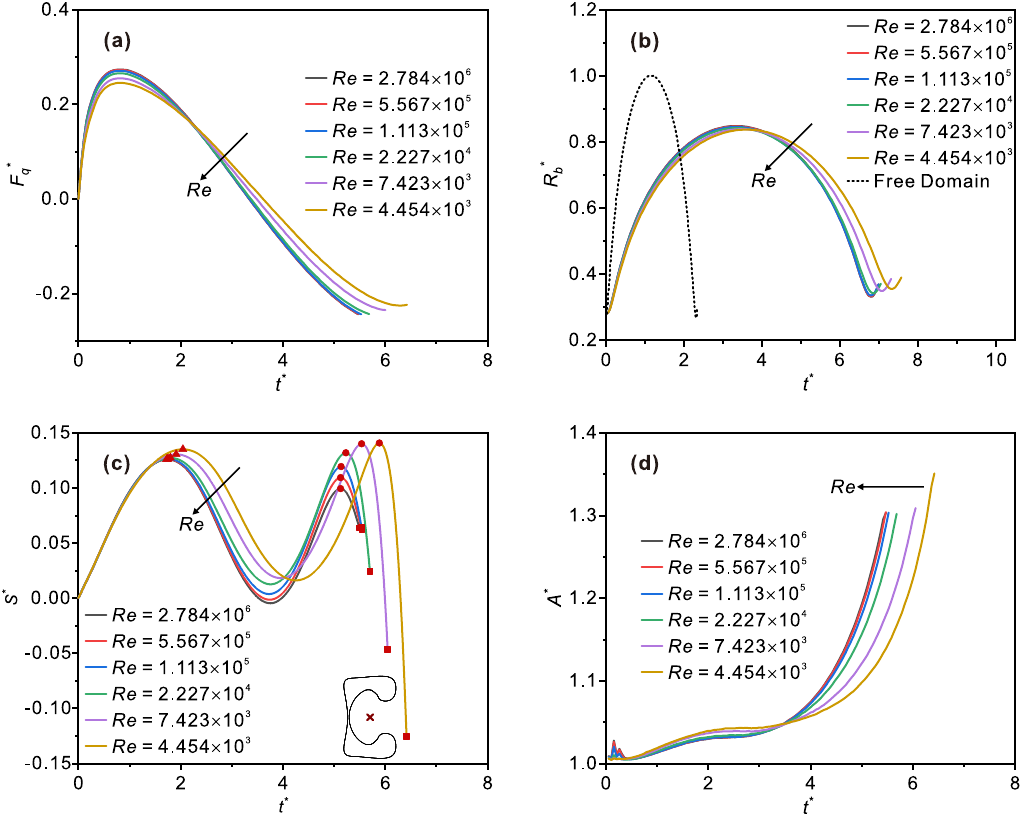}
  \caption{Effect of the Reynolds number $Re$ on the bubble shape with $R_t^*$ = 0.802, $L_t^*$ = 8.556, and $U^*$ = 0.088: (a) Variation of mass flow rate ratio $F_q^*$; (b) Variation of bubble size $R_b^*$; (c) Variation of bubble skewness $S^*$; (d) Variation of bubble deformation $A^*$.}\label{fig:14}
\end{figure}

To analyze the effect of liquid viscosity on bubble evolution, the Reynolds number is defined
\begin{equation}\label{eq:25}
	Re=\frac{2{{\rho }_{l}}{{v}_{m}}{{R}_{t}}}{{{\mu }_{l}}}
\end{equation}
The influence of $Re$ on the mass flow rate ratio $F_q^*$ is shown in Figure \ref{fig:14}(a). As the Reynolds number $Re$ decreases, the maximum value of $F_q^*$ first remains almost constant and then decreases, and the magnitude of $F_q^*$ at the moment when the bubble's two side interfaces meet first remains almost constant and then decreases (note that at this time $F_q^* < 0$). This is because when $Re$ is large (indicating low liquid viscosity), the viscous forces are relatively small compared to the inertial forces, making the influence of viscosity on bubble evolution negligible. In such cases, variation in $Re$ has little effect on $F_q^*$. However, when $Re$ becomes sufficiently small (${Re} < 7.423\times10^3$), the impact of viscous forces on bubble evolution becomes significant. In this scenario, the decrease in $Re$ enhances the inhibitory effect of viscosity on the flow, resulting in a decrease in the expansion and contraction rates of the bubble and a reduction in the bubble's influence on the outlet flow rate. Figure \ref{fig:14}(b) further illustrates the impact of $Re$ on the bubble size $R_b^*$. As $Re$ decreases, the influence of the liquid viscosity becomes more pronounced, leading to the bubble's lifetime first remaining almost constant and then increasing. When $Re$ is sufficiently small, the decrease in $Re$ strengthens the inhibitory effect of viscous forces on the bubble, slowing down the expansion and contraction speeds of the bubble, and consequently prolonging its lifetime. Unlike the results in stagnant-water tubes \cite{Wang2024WallConf}, as $Re$ decreases, the maximum volume of the bubbles in flowing tubes does not increase but decreases. This is because in stagnant-water tubes, an increase in the viscous force leads to an extended bubble expansion process, resulting in an increase in the maximum volume of the bubble. However, in flowing tubes, an increase in viscous forces leads to increased flow resistance within the tube, reducing the flow velocity through the gap between the bubble and the tube wall. With the inlet flow rate remaining constant, the reduction in flow velocity within the gap causes the liquid to radially compress the bubble to increase the flow area, thereby restricting the bubble's expansion and resulting in a decrease in the maximum volume of the bubble.

The effect of the Reynolds number $Re$ on the bubble skewness $S^*$ is shown in Figure \ref{fig:14}(c). As $Re$ decreases, both the peak value of $S^*$ during the expansion (the first peak, as shown by the triangle symbol in Figure \ref{fig:14}(c)) and during the contraction (the second peak, as shown by the circular symbol in Figure \ref{fig:14}(c)) increase. The value at the moment when the two sides of the bubble interface meet (as shown by the square symbol in Figure \ref{fig:14}(c)) first decreases then increases (note that when ${Re} < 7.423\times10^3$, $S^* < 0$). This is because during the bubble expansion process, as $Re$ decreases, the enhancement of liquid viscosity prolongs the expansion process on the right-hand side of the bubble. The increase in the expansion distance of the right half of the bubble results in an increase in its length, thereby increasing the difference in lengths between the two parts of the bubble. During the bubble's contraction process, the strengthening of the viscous forces prolongs the contraction, allowing the liquid jet to fully develop. This results in an increase in the jet diameter, causing the right half of the annular bubble to thin out and the length of the right half to increase. As a result, the difference in length between the left and right halves of the bubble increases. When the bubble's two side interfaces meet, the increased viscosity causes the flow entering the gap between the bubble and the tube wall to deflect under the influence of the right-hand-side liquid jet. This deflection pushes the right-hand side of the bubble closer to the axis. As a result, the length of the right half of the bubble decreases, reducing the difference in lengths between the left and right parts of the bubble. When $Re$ is very small (${Re} < 7.423\times10^3$), the length of the bubble's right half decreases to less than that of the left half, resulting in $S^* < 0$ (as shown in the inset of Figure \ref{fig:14}(c)), this indicates an increased difference in the lengths of the two sides of the bubble. Under different Reynolds numbers $Re$, the asymmetrical behavior of the bubble on both sides affects the degree of bubble deformation, as shown in Figure \ref{fig:14}(d). As $Re$ decreases, the degree of deformation first remains constant and then increases. This is because, with decreasing $Re$, the strengthening of viscous effects prolongs the evolution process of the bubble. The right-hand-side liquid jet is allowed to fully develop, leading to an increase in its diameter. This causes an increase in the deformation degree of the bubble before the two side interfaces meet.

\begin{figure}
  \centering
  \includegraphics[scale=0.9]{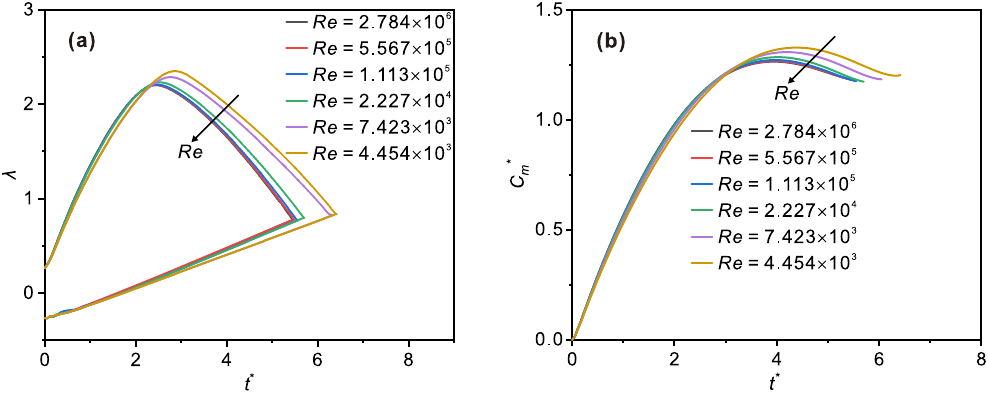}
  \caption{Effect of the Reynolds number $Re$ on the bubble position with $R_t^*$ = 0.802, $L_t^*$ = 8.556, and $U^*$ = 0.088: (a) Variation of interface positions on the two sides of the bubble $\lambda$; (b) Variation of bubble's center of mass position $C_m^*$. }\label{fig:15}
\end{figure}

The influence of $Re$ on the positions of the two bubble side interfaces $\lambda$ and the bubble's center of mass position $C_m^*$ also shows a trend of first remaining constant and then significantly changing, as shown in Figures \ref{fig:15}(a)-(b). When the Reynolds number $Re$ is sufficiently small, a decrease in $Re$ strengthens the inhibitory effect of viscosity on the flow. This results in a decrease in the velocity of the left bubble interface and a decrease in both the expansion and contraction speeds of the right interface. The decrease in $Re$ prolongs the expansion process, leading to an increase in the expansion distance of the right interface and causing the meeting position of the two interfaces to shift toward the outlet of the tube. Additionally, when the Reynolds number $Re$ is sufficiently small, a decrease in $Re$ leads to an increase in the final movement distance of the bubble's center of mass. This is because the enhanced viscous effect prolongs the evolution time of the bubble, resulting in an increased migration distance under the drive of the inflow.

\section{Conclusions}\label{sec:4}
In this study, we conduct a numerical simulation on the dynamics of a cavitation bubble in a flowing tube, focusing on the asymmetric behavior of the bubble under the combined effects of inflow and tube wall. During the evolution of the bubble, significant asymmetry is observed in both the axial positions and shapes of the bubble interfaces. Factors such as tube diameter, tube length, inflow velocity, and liquid viscosity have a significant effect on the behavior of bubble evolution. As the tube diameter increases, the effects of inflow and tube wall on the bubble both decrease. This results in an increase in the maximum volume of the bubble, a decrease in bubble lifetime, a decrease in deformation degree, a decrease in migration distance, and initially a decrease followed by an increase in the bubble skewness (degree of asymmetry of lengths on both sides). Increasing tube length enhances the restrictive effect of tube wall on the bubble, resulting in prolonged bubble lifetime, increased deformation degree, increased migration distance, and increased bubble skewness. The degree of bubble movement toward the outlet of the tube increases with the increase in inflow velocity. This results in a decrease in the maximum volume of the bubble, a decrease in bubble lifetime, a decrease in deformation degree, an increase in migration distance, and a trend in the bubble skewness where it decreases initially, then increases, and finally decreases again. The increase in liquid viscosity strengthens the viscous effect of tube wall, restricting the evolution of the bubble. This leads to an increase in bubble lifetime, an increase in deformation degree, an increase in migration distance, and an initial decrease followed by a later increase in bubble skewness. Additionally, the evolution of the bubble has a significant effect on the flow in the tube. The expansion of the bubble leads to an increase in the mass flow rate at the tube outlet, while the contraction of the bubble causes a decrease in the mass flow rate at the outlet, and even the occurrence of backflow.

\section*{Acknowledgements}
This work is supported by the National Natural Science Foundation of China (Grant nos.\ 51920105010 and 51921004).

\section*{Data Availability}
The data that support the findings of this study are available from the corresponding author upon reasonable request.
\section*{References}
\bibliography{cavitationFlowBubble}

\begin{thebibliography}{54}%
\makeatletter
\providecommand \@ifxundefined [1]{%
 \@ifx{#1\undefined}
}%
\providecommand \@ifnum [1]{%
 \ifnum #1\expandafter \@firstoftwo
 \else \expandafter \@secondoftwo
 \fi
}%
\providecommand \@ifx [1]{%
 \ifx #1\expandafter \@firstoftwo
 \else \expandafter \@secondoftwo
 \fi
}%
\providecommand \natexlab [1]{#1}%
\providecommand \enquote  [1]{``#1''}%
\providecommand \bibnamefont  [1]{#1}%
\providecommand \bibfnamefont [1]{#1}%
\providecommand \citenamefont [1]{#1}%
\providecommand \href@noop [0]{\@secondoftwo}%
\providecommand \href [0]{\begingroup \@sanitize@url \@href}%
\providecommand \@href[1]{\@@startlink{#1}\@@href}%
\providecommand \@@href[1]{\endgroup#1\@@endlink}%
\providecommand \@sanitize@url [0]{\catcode `\\12\catcode `\$12\catcode
  `\&12\catcode `\#12\catcode `\^12\catcode `\_12\catcode `\%12\relax}%
\providecommand \@@startlink[1]{}%
\providecommand \@@endlink[0]{}%
\providecommand \url  [0]{\begingroup\@sanitize@url \@url }%
\providecommand \@url [1]{\endgroup\@href {#1}{\urlprefix }}%
\providecommand \urlprefix  [0]{URL }%
\providecommand \Eprint [0]{\href }%
\providecommand \doibase [0]{http://dx.doi.org/}%
\providecommand \selectlanguage [0]{\@gobble}%
\providecommand \bibinfo  [0]{\@secondoftwo}%
\providecommand \bibfield  [0]{\@secondoftwo}%
\providecommand \translation [1]{[#1]}%
\providecommand \BibitemOpen [0]{}%
\providecommand \bibitemStop [0]{}%
\providecommand \bibitemNoStop [0]{.\EOS\space}%
\providecommand \EOS [0]{\spacefactor3000\relax}%
\providecommand \BibitemShut  [1]{\csname bibitem#1\endcsname}%
\let\auto@bib@innerbib\@empty
\bibitem [{\citenamefont {Bilus}(2020)}]{Bilus2020}%
  \BibitemOpen
  \bibfield  {author} {\bibinfo {author} {\bibfnamefont {L.}~\bibnamefont
  {Bilus}, \bibfnamefont {Matevz~Lesnik}},\ }\bibfield  {title} {\enquote
  {\bibinfo {title} {Numerical prediction of various cavitation erosion
  mechanisms},}\ }\href {\doibase 10.1115/1.4045365} {\bibfield  {journal}
  {\bibinfo  {journal} {Journal of Fluids Engineering-Transactions of the
  ASME}\ }\textbf {\bibinfo {volume} {142}},\ \bibinfo {pages} {041402}
  (\bibinfo {year} {2020})}\BibitemShut {NoStop}%
\bibitem [{\citenamefont {Du}\ and\ \citenamefont {Chen}(2021)}]{Du2021}%
  \BibitemOpen
  \bibfield  {author} {\bibinfo {author} {\bibfnamefont {J.}~\bibnamefont
  {Du}}\ and\ \bibinfo {author} {\bibfnamefont {F.}~\bibnamefont {Chen}},\
  }\bibfield  {title} {\enquote {\bibinfo {title} {Cavitation dynamics and flow
  aggressiveness in ultrasonic cavitation erosion},}\ }\href {\doibase
  10.1016/j.ijmecsci.2021.106545} {\bibfield  {journal} {\bibinfo  {journal}
  {International Journal of Mechanical Sciences}\ }\textbf {\bibinfo {volume}
  {204}},\ \bibinfo {pages} {106545} (\bibinfo {year} {2021})}\BibitemShut
  {NoStop}%
\bibitem [{\citenamefont {Li}\ \emph {et~al.}(2023{\natexlab{a}})\citenamefont
  {Li}, \citenamefont {Niu}, \citenamefont {Wei}, \citenamefont {Manickam},
  \citenamefont {Sun},\ and\ \citenamefont {Zhu}}]{LiLINMIN2023}%
  \BibitemOpen
  \bibfield  {author} {\bibinfo {author} {\bibfnamefont {L.}~\bibnamefont
  {Li}}, \bibinfo {author} {\bibfnamefont {Y.}~\bibnamefont {Niu}}, \bibinfo
  {author} {\bibfnamefont {G.}~\bibnamefont {Wei}}, \bibinfo {author}
  {\bibfnamefont {S.}~\bibnamefont {Manickam}}, \bibinfo {author}
  {\bibfnamefont {X.}~\bibnamefont {Sun}}, \ and\ \bibinfo {author}
  {\bibfnamefont {Z.}~\bibnamefont {Zhu}},\ }\bibfield  {title} {\enquote
  {\bibinfo {title} {Investigation of cavitation noise using
  eulerian-lagrangian multiscale modeling},}\ }\href {\doibase
  10.2139/ssrn.4426924} {\bibfield  {journal} {\bibinfo  {journal} {Ultrasonics
  Sonochemistry}\ }\textbf {\bibinfo {volume} {97}},\ \bibinfo {pages} {106446}
  (\bibinfo {year} {2023}{\natexlab{a}})}\BibitemShut {NoStop}%
\bibitem [{\citenamefont {Si}\ \emph {et~al.}(2023)\citenamefont {Si},
  \citenamefont {Ali}, \citenamefont {Liao}, \citenamefont {Yuan},
  \citenamefont {Gu}, \citenamefont {Yuan},\ and\ \citenamefont
  {Bois}}]{Si2023}%
  \BibitemOpen
  \bibfield  {author} {\bibinfo {author} {\bibfnamefont {Q.~R.}\ \bibnamefont
  {Si}}, \bibinfo {author} {\bibfnamefont {A.}~\bibnamefont {Ali}}, \bibinfo
  {author} {\bibfnamefont {M.~Q.}\ \bibnamefont {Liao}}, \bibinfo {author}
  {\bibfnamefont {J.~P.}\ \bibnamefont {Yuan}}, \bibinfo {author}
  {\bibfnamefont {Y.~Y.}\ \bibnamefont {Gu}}, \bibinfo {author} {\bibfnamefont
  {S.~Q.}\ \bibnamefont {Yuan}}, \ and\ \bibinfo {author} {\bibfnamefont
  {G.}~\bibnamefont {Bois}},\ }\bibfield  {title} {\enquote {\bibinfo {title}
  {Assessment of cavitation noise in a centrifugal pump using acoustic finite
  element method and spherical cavity radiation theory},}\ }\href {\doibase
  10.1080/19942060.2023.2173302} {\bibfield  {journal} {\bibinfo  {journal}
  {Engineering Applications of Computational Fluid Mechanics}\ }\textbf
  {\bibinfo {volume} {17}},\ \bibinfo {pages} {2173302} (\bibinfo {year}
  {2023})}\BibitemShut {NoStop}%
\bibitem [{\citenamefont {Nanzai}\ \emph {et~al.}(2023)\citenamefont {Nanzai},
  \citenamefont {Mochizuki}, \citenamefont {Wakikawa}, \citenamefont {Masuda},
  \citenamefont {Oshio},\ and\ \citenamefont {Yagishita}}]{Nanzai2023}%
  \BibitemOpen
  \bibfield  {author} {\bibinfo {author} {\bibfnamefont {B.}~\bibnamefont
  {Nanzai}}, \bibinfo {author} {\bibfnamefont {A.}~\bibnamefont {Mochizuki}},
  \bibinfo {author} {\bibfnamefont {Y.}~\bibnamefont {Wakikawa}}, \bibinfo
  {author} {\bibfnamefont {Y.}~\bibnamefont {Masuda}}, \bibinfo {author}
  {\bibfnamefont {T.}~\bibnamefont {Oshio}}, \ and\ \bibinfo {author}
  {\bibfnamefont {K.}~\bibnamefont {Yagishita}},\ }\bibfield  {title} {\enquote
  {\bibinfo {title} {Sonoluminescence intensity and ultrasonic cavitation
  temperature in organic solvents: Effects of generated radicals},}\ }\href
  {\doibase 10.1016/j.ultsonch.2023.106357} {\bibfield  {journal} {\bibinfo
  {journal} {Ultrasonics Sonochemistry}\ }\textbf {\bibinfo {volume} {95}},\
  \bibinfo {pages} {106357} (\bibinfo {year} {2023})}\BibitemShut {NoStop}%
\bibitem [{\citenamefont {Yusof}\ \emph {et~al.}(2022)\citenamefont {Yusof},
  \citenamefont {Anandan}, \citenamefont {Sivashanmugam}, \citenamefont
  {Flores},\ and\ \citenamefont {Ashokkumar}}]{Yusof2022}%
  \BibitemOpen
  \bibfield  {author} {\bibinfo {author} {\bibfnamefont {N.~S.~M.}\
  \bibnamefont {Yusof}}, \bibinfo {author} {\bibfnamefont {S.}~\bibnamefont
  {Anandan}}, \bibinfo {author} {\bibfnamefont {P.}~\bibnamefont
  {Sivashanmugam}}, \bibinfo {author} {\bibfnamefont {E.~M.}\ \bibnamefont
  {Flores}}, \ and\ \bibinfo {author} {\bibfnamefont {M.}~\bibnamefont
  {Ashokkumar}},\ }\bibfield  {title} {\enquote {\bibinfo {title} {A
  correlation between cavitation bubble temperature, sonoluminescence and
  interfacial chemistry--a minireview},}\ }\href {\doibase
  10.1016/j.ultsonch.2022.105988} {\bibfield  {journal} {\bibinfo  {journal}
  {Ultrasonics Sonochemistry}\ }\textbf {\bibinfo {volume} {85}},\ \bibinfo
  {pages} {105988} (\bibinfo {year} {2022})}\BibitemShut {NoStop}%
\bibitem [{\citenamefont {Ebrahimi}\ \emph {et~al.}(2021)\citenamefont
  {Ebrahimi}, \citenamefont {Razaghian}, \citenamefont {Tootian},\ and\
  \citenamefont {Seif}}]{Ebrahimi2021}%
  \BibitemOpen
  \bibfield  {author} {\bibinfo {author} {\bibfnamefont {A.}~\bibnamefont
  {Ebrahimi}}, \bibinfo {author} {\bibfnamefont {A.}~\bibnamefont {Razaghian}},
  \bibinfo {author} {\bibfnamefont {A.}~\bibnamefont {Tootian}}, \ and\
  \bibinfo {author} {\bibfnamefont {M.}~\bibnamefont {Seif}},\ }\bibfield
  {title} {\enquote {\bibinfo {title} {An experimental investigation of
  hydrodynamic performance, cavitation, and noise of a normal skew {B}-series
  marine propeller in the cavitation tunnel},}\ }\href {\doibase
  10.1016/j.oceaneng.2021.109739} {\bibfield  {journal} {\bibinfo  {journal}
  {Ocean Engineering}\ }\textbf {\bibinfo {volume} {238}},\ \bibinfo {pages}
  {109739} (\bibinfo {year} {2021})}\BibitemShut {NoStop}%
\bibitem [{\citenamefont {Jia}\ \emph {et~al.}(2023)\citenamefont {Jia},
  \citenamefont {Zhang}, \citenamefont {Lv},\ and\ \citenamefont
  {Zhu}}]{JiaX2023}%
  \BibitemOpen
  \bibfield  {author} {\bibinfo {author} {\bibfnamefont {X.~Q.}\ \bibnamefont
  {Jia}}, \bibinfo {author} {\bibfnamefont {Y.}~\bibnamefont {Zhang}}, \bibinfo
  {author} {\bibfnamefont {H.}~\bibnamefont {Lv}}, \ and\ \bibinfo {author}
  {\bibfnamefont {Z.~C.}\ \bibnamefont {Zhu}},\ }\bibfield  {title} {\enquote
  {\bibinfo {title} {Study on external performance and internal flow
  characteristics in a centrifugal pump under different degrees of
  cavitation},}\ }\href {\doibase 10.1063/5.0133377} {\bibfield  {journal}
  {\bibinfo  {journal} {Physics of Fluids}\ }\textbf {\bibinfo {volume} {35}},\
  \bibinfo {pages} {014104} (\bibinfo {year} {2023})}\BibitemShut {NoStop}%
\bibitem [{\citenamefont {Wang}, \citenamefont {Cheng},\ and\ \citenamefont
  {Ji}(2022)}]{Wang2022Num}%
  \BibitemOpen
  \bibfield  {author} {\bibinfo {author} {\bibfnamefont {Z.}~\bibnamefont
  {Wang}}, \bibinfo {author} {\bibfnamefont {H.}~\bibnamefont {Cheng}}, \ and\
  \bibinfo {author} {\bibfnamefont {B.}~\bibnamefont {Ji}},\ }\bibfield
  {title} {\enquote {\bibinfo {title} {Numerical prediction of cavitation
  erosion risk in an axisymmetric nozzle using a multi-scale approach},}\
  }\href {\doibase 10.1063/5.0095833} {\bibfield  {journal} {\bibinfo
  {journal} {Physics of Fluids}\ }\textbf {\bibinfo {volume} {34}},\ \bibinfo
  {pages} {062112} (\bibinfo {year} {2022})}\BibitemShut {NoStop}%
\bibitem [{\citenamefont {Nikolaev}\ \emph {et~al.}(2018)\citenamefont
  {Nikolaev}, \citenamefont {Gopin}, \citenamefont {Severin}, \citenamefont
  {Rudin}, \citenamefont {Mironov},\ and\ \citenamefont
  {Dezhkunov}}]{Nikolaev2018}%
  \BibitemOpen
  \bibfield  {author} {\bibinfo {author} {\bibfnamefont {A.}~\bibnamefont
  {Nikolaev}}, \bibinfo {author} {\bibfnamefont {A.}~\bibnamefont {Gopin}},
  \bibinfo {author} {\bibfnamefont {A.}~\bibnamefont {Severin}}, \bibinfo
  {author} {\bibfnamefont {V.}~\bibnamefont {Rudin}}, \bibinfo {author}
  {\bibfnamefont {M.}~\bibnamefont {Mironov}}, \ and\ \bibinfo {author}
  {\bibfnamefont {N.}~\bibnamefont {Dezhkunov}},\ }\bibfield  {title} {\enquote
  {\bibinfo {title} {Ultrasonic synthesis of hydroxyapatite in non-cavitation
  and cavitation modes},}\ }\href {\doibase 10.1016/j.ultsonch.2018.02.047}
  {\bibfield  {journal} {\bibinfo  {journal} {Ultrasonics sonochemistry}\
  }\textbf {\bibinfo {volume} {44}},\ \bibinfo {pages} {390--397} (\bibinfo
  {year} {2018})}\BibitemShut {NoStop}%
\bibitem [{\citenamefont {Rezk}\ \emph {et~al.}(2021)\citenamefont {Rezk},
  \citenamefont {Ahmed}, \citenamefont {Ramesan},\ and\ \citenamefont
  {Yeo}}]{Rezk2021}%
  \BibitemOpen
  \bibfield  {author} {\bibinfo {author} {\bibfnamefont {A.~R.}\ \bibnamefont
  {Rezk}}, \bibinfo {author} {\bibfnamefont {H.}~\bibnamefont {Ahmed}},
  \bibinfo {author} {\bibfnamefont {S.}~\bibnamefont {Ramesan}}, \ and\
  \bibinfo {author} {\bibfnamefont {L.~Y.}\ \bibnamefont {Yeo}},\ }\bibfield
  {title} {\enquote {\bibinfo {title} {High frequency sonoprocessing: A new
  field of cavitation-free acoustic materials synthesis, processing, and
  manipulation},}\ }\href {\doibase 10.1002/advs.202001983} {\bibfield
  {journal} {\bibinfo  {journal} {Advanced Science}\ }\textbf {\bibinfo
  {volume} {8}},\ \bibinfo {pages} {2001983} (\bibinfo {year}
  {2021})}\BibitemShut {NoStop}%
\bibitem [{\citenamefont {Xu}, \citenamefont {Zeiger},\ and\ \citenamefont
  {Suslick}(2013)}]{Xu2013}%
  \BibitemOpen
  \bibfield  {author} {\bibinfo {author} {\bibfnamefont {H.~X.}\ \bibnamefont
  {Xu}}, \bibinfo {author} {\bibfnamefont {B.~W.}\ \bibnamefont {Zeiger}}, \
  and\ \bibinfo {author} {\bibfnamefont {K.~S.}\ \bibnamefont {Suslick}},\
  }\bibfield  {title} {\enquote {\bibinfo {title} {Sonochemical synthesis of
  nanomaterials},}\ }\href {\doibase 10.1039/C2CS35282F} {\bibfield  {journal}
  {\bibinfo  {journal} {Chemical Society Reviews}\ }\textbf {\bibinfo {volume}
  {42}},\ \bibinfo {pages} {2555--2567} (\bibinfo {year} {2013})}\BibitemShut
  {NoStop}%
\bibitem [{\citenamefont {Brennen}(2015)}]{Brennen2015}%
  \BibitemOpen
  \bibfield  {author} {\bibinfo {author} {\bibfnamefont {C.~E.}\ \bibnamefont
  {Brennen}},\ }\bibfield  {title} {\enquote {\bibinfo {title} {Cavitation in
  medicine},}\ }\href {\doibase 10.1098/rsfs.2015.0022} {\bibfield  {journal}
  {\bibinfo  {journal} {Interface Focus}\ }\textbf {\bibinfo {volume} {5}},\
  \bibinfo {pages} {20150022} (\bibinfo {year} {2015})}\BibitemShut {NoStop}%
\bibitem [{\citenamefont {Chen}\ \emph {et~al.}(2011)\citenamefont {Chen},
  \citenamefont {Kreider}, \citenamefont {Brayman}, \citenamefont {Bailey},\
  and\ \citenamefont {Matula}}]{Chen2011}%
  \BibitemOpen
  \bibfield  {author} {\bibinfo {author} {\bibfnamefont {H.}~\bibnamefont
  {Chen}}, \bibinfo {author} {\bibfnamefont {W.}~\bibnamefont {Kreider}},
  \bibinfo {author} {\bibfnamefont {A.~A.}\ \bibnamefont {Brayman}}, \bibinfo
  {author} {\bibfnamefont {M.~R.}\ \bibnamefont {Bailey}}, \ and\ \bibinfo
  {author} {\bibfnamefont {T.~J.}\ \bibnamefont {Matula}},\ }\bibfield  {title}
  {\enquote {\bibinfo {title} {Blood vessel deformations on microsecond time
  scales by ultrasonic cavitation},}\ }\href {\doibase
  10.1103/PhysRevLett.106.034301} {\bibfield  {journal} {\bibinfo  {journal}
  {Physical Review Letters}\ }\textbf {\bibinfo {volume} {106}},\ \bibinfo
  {pages} {034301} (\bibinfo {year} {2011})}\BibitemShut {NoStop}%
\bibitem [{\citenamefont {Hosny}\ \emph {et~al.}(2013)\citenamefont {Hosny},
  \citenamefont {Mohamedi}, \citenamefont {Rademeyer}, \citenamefont {Owen},
  \citenamefont {Wu}, \citenamefont {Tang}, \citenamefont {Eckersley},
  \citenamefont {Stride},\ and\ \citenamefont {Kuimova}}]{Hosny2013}%
  \BibitemOpen
  \bibfield  {author} {\bibinfo {author} {\bibfnamefont {N.~A.}\ \bibnamefont
  {Hosny}}, \bibinfo {author} {\bibfnamefont {G.}~\bibnamefont {Mohamedi}},
  \bibinfo {author} {\bibfnamefont {P.}~\bibnamefont {Rademeyer}}, \bibinfo
  {author} {\bibfnamefont {J.}~\bibnamefont {Owen}}, \bibinfo {author}
  {\bibfnamefont {Y.~L.}\ \bibnamefont {Wu}}, \bibinfo {author} {\bibfnamefont
  {M.~X.}\ \bibnamefont {Tang}}, \bibinfo {author} {\bibfnamefont {R.~J.}\
  \bibnamefont {Eckersley}}, \bibinfo {author} {\bibfnamefont {E.}~\bibnamefont
  {Stride}}, \ and\ \bibinfo {author} {\bibfnamefont {M.~K.}\ \bibnamefont
  {Kuimova}},\ }\bibfield  {title} {\enquote {\bibinfo {title} {Mapping
  microbubble viscosity using fluorescence lifetime imaging of molecular
  rotors},}\ }\href {\doibase 10.1073/pnas.1301479110} {\bibfield  {journal}
  {\bibinfo  {journal} {Proceedings of the National Academy of Sciences of the
  United States of America}\ }\textbf {\bibinfo {volume} {110}},\ \bibinfo
  {pages} {9225--9230} (\bibinfo {year} {2013})}\BibitemShut {NoStop}%
\bibitem [{\citenamefont {Chahine}\ \emph {et~al.}(2016)\citenamefont
  {Chahine}, \citenamefont {Kapahi}, \citenamefont {Choi},\ and\ \citenamefont
  {Hsiao}}]{Chahine2016}%
  \BibitemOpen
  \bibfield  {author} {\bibinfo {author} {\bibfnamefont {G.~L.}\ \bibnamefont
  {Chahine}}, \bibinfo {author} {\bibfnamefont {A.}~\bibnamefont {Kapahi}},
  \bibinfo {author} {\bibfnamefont {J.~K.}\ \bibnamefont {Choi}}, \ and\
  \bibinfo {author} {\bibfnamefont {C.~T.}\ \bibnamefont {Hsiao}},\ }\bibfield
  {title} {\enquote {\bibinfo {title} {Modeling of surface cleaning by
  cavitation bubble dynamics and collapse},}\ }\href {\doibase
  10.1016/j.ultsonch.2015.04.026} {\bibfield  {journal} {\bibinfo  {journal}
  {Ultrasonics Sonochemistry}\ }\textbf {\bibinfo {volume} {29}},\ \bibinfo
  {pages} {528--549} (\bibinfo {year} {2016})}\BibitemShut {NoStop}%
\bibitem [{\citenamefont {Li}\ \emph {et~al.}(2022)\citenamefont {Li},
  \citenamefont {Wang}, \citenamefont {Liao}, \citenamefont {Ueda},
  \citenamefont {Yoshikawa},\ and\ \citenamefont {Zhang}}]{LiPan2022Bubble}%
  \BibitemOpen
  \bibfield  {author} {\bibinfo {author} {\bibfnamefont {P.}~\bibnamefont
  {Li}}, \bibinfo {author} {\bibfnamefont {J.}~\bibnamefont {Wang}}, \bibinfo
  {author} {\bibfnamefont {Z.}~\bibnamefont {Liao}}, \bibinfo {author}
  {\bibfnamefont {Y.}~\bibnamefont {Ueda}}, \bibinfo {author} {\bibfnamefont
  {K.}~\bibnamefont {Yoshikawa}}, \ and\ \bibinfo {author} {\bibfnamefont
  {G.}~\bibnamefont {Zhang}},\ }\bibfield  {title} {\enquote {\bibinfo {title}
  {Microbubbles for effective cleaning of metal surfaces without chemical
  agents},}\ }\href {\doibase 10.1021/acs.langmuir.1c02769} {\bibfield
  {journal} {\bibinfo  {journal} {Langmuir}\ }\textbf {\bibinfo {volume}
  {38}},\ \bibinfo {pages} {769--776} (\bibinfo {year} {2022})}\BibitemShut
  {NoStop}%
\bibitem [{\citenamefont {Zhong}\ \emph {et~al.}(2022)\citenamefont {Zhong},
  \citenamefont {Dong}, \citenamefont {Liu}, \citenamefont {Meng},
  \citenamefont {Li},\ and\ \citenamefont {Pan}}]{Zhong2022}%
  \BibitemOpen
  \bibfield  {author} {\bibinfo {author} {\bibfnamefont {X.}~\bibnamefont
  {Zhong}}, \bibinfo {author} {\bibfnamefont {J.}~\bibnamefont {Dong}},
  \bibinfo {author} {\bibfnamefont {M.}~\bibnamefont {Liu}}, \bibinfo {author}
  {\bibfnamefont {R.}~\bibnamefont {Meng}}, \bibinfo {author} {\bibfnamefont
  {S.}~\bibnamefont {Li}}, \ and\ \bibinfo {author} {\bibfnamefont
  {X.}~\bibnamefont {Pan}},\ }\bibfield  {title} {\enquote {\bibinfo {title}
  {Experimental study on ship fouling cleaning by ultrasonic-enhanced submerged
  cavitation jet: A preliminary study},}\ }\href {\doibase
  10.1016/j.oceaneng.2022.111844} {\bibfield  {journal} {\bibinfo  {journal}
  {Ocean Engineering}\ }\textbf {\bibinfo {volume} {258}},\ \bibinfo {pages}
  {111844} (\bibinfo {year} {2022})}\BibitemShut {NoStop}%
\bibitem [{\citenamefont {Rayleigh}(1917)}]{Rayleigh1917}%
  \BibitemOpen
  \bibfield  {author} {\bibinfo {author} {\bibfnamefont {L.}~\bibnamefont
  {Rayleigh}},\ }\bibfield  {title} {\enquote {\bibinfo {title} {{VIII.} on the
  pressure developed in a liquid during the collapse of a spherical cavity},}\
  }\href {\doibase 10.1080/14786440808635681} {\bibfield  {journal} {\bibinfo
  {journal} {The London, Edinburgh, and Dublin Philosophical Magazine and
  Journal of Science}\ }\textbf {\bibinfo {volume} {34}},\ \bibinfo {pages}
  {94--98} (\bibinfo {year} {1917})}\BibitemShut {NoStop}%
\bibitem [{\citenamefont {Plesset}(1949)}]{Plesset1949}%
  \BibitemOpen
  \bibfield  {author} {\bibinfo {author} {\bibfnamefont {M.~S.}\ \bibnamefont
  {Plesset}},\ }\bibfield  {title} {\enquote {\bibinfo {title} {The dynamics of
  cavitation bubbles},}\ }\href {\doibase 10.1115/1.4009975} {\bibfield
  {journal} {\bibinfo  {journal} {Journal of Applied Mechanics}\ }\textbf
  {\bibinfo {volume} {16}},\ \bibinfo {pages} {277--282} (\bibinfo {year}
  {1949})}\BibitemShut {NoStop}%
\bibitem [{\citenamefont {Keller}\ and\ \citenamefont
  {Miksis}(1980)}]{Keller1980}%
  \BibitemOpen
  \bibfield  {author} {\bibinfo {author} {\bibfnamefont {J.~B.}\ \bibnamefont
  {Keller}}\ and\ \bibinfo {author} {\bibfnamefont {M.}~\bibnamefont
  {Miksis}},\ }\bibfield  {title} {\enquote {\bibinfo {title} {Bubble
  oscillations of large amplitude},}\ }\href {\doibase 10.1121/1.384720}
  {\bibfield  {journal} {\bibinfo  {journal} {The Journal of the Acoustical
  Society of America}\ }\textbf {\bibinfo {volume} {68}},\ \bibinfo {pages}
  {628--633} (\bibinfo {year} {1980})}\BibitemShut {NoStop}%
\bibitem [{\citenamefont {Zhang}\ \emph {et~al.}(2023)\citenamefont {Zhang},
  \citenamefont {Li}, \citenamefont {Cui}, \citenamefont {Li},\ and\
  \citenamefont {Liu}}]{Zhang2023A}%
  \BibitemOpen
  \bibfield  {author} {\bibinfo {author} {\bibfnamefont {A.-M.}\ \bibnamefont
  {Zhang}}, \bibinfo {author} {\bibfnamefont {S.-M.}\ \bibnamefont {Li}},
  \bibinfo {author} {\bibfnamefont {P.}~\bibnamefont {Cui}}, \bibinfo {author}
  {\bibfnamefont {S.}~\bibnamefont {Li}}, \ and\ \bibinfo {author}
  {\bibfnamefont {Y.-L.}\ \bibnamefont {Liu}},\ }\bibfield  {title} {\enquote
  {\bibinfo {title} {A unified theory for bubble dynamics},}\ }\href {\doibase
  10.1063/5.0145415} {\bibfield  {journal} {\bibinfo  {journal} {Physics of
  Fluids}\ }\textbf {\bibinfo {volume} {35}},\ \bibinfo {pages} {033323}
  (\bibinfo {year} {2023})}\BibitemShut {NoStop}%
\bibitem [{\citenamefont {Wang}\ \emph {et~al.}(2020)\citenamefont {Wang},
  \citenamefont {Mahmud}, \citenamefont {Cui}, \citenamefont {Smith},\ and\
  \citenamefont {Walmsley}}]{Wang2022}%
  \BibitemOpen
  \bibfield  {author} {\bibinfo {author} {\bibfnamefont {Q.~X.}\ \bibnamefont
  {Wang}}, \bibinfo {author} {\bibfnamefont {M.}~\bibnamefont {Mahmud}},
  \bibinfo {author} {\bibfnamefont {J.}~\bibnamefont {Cui}}, \bibinfo {author}
  {\bibfnamefont {W.~R.}\ \bibnamefont {Smith}}, \ and\ \bibinfo {author}
  {\bibfnamefont {A.~D.}\ \bibnamefont {Walmsley}},\ }\bibfield  {title}
  {\enquote {\bibinfo {title} {Numerical investigation of bubble dynamics at a
  corner},}\ }\href {\doibase 10.1063/1.5140740} {\bibfield  {journal}
  {\bibinfo  {journal} {Physics of Fluids}\ }\textbf {\bibinfo {volume} {32}},\
  \bibinfo {pages} {053306} (\bibinfo {year} {2020})}\BibitemShut {NoStop}%
\bibitem [{\citenamefont {Xu}\ \emph {et~al.}(2023)\citenamefont {Xu},
  \citenamefont {Li}, \citenamefont {Ren}, \citenamefont {Liu},\ and\
  \citenamefont {Zuo}}]{Xu2023}%
  \BibitemOpen
  \bibfield  {author} {\bibinfo {author} {\bibfnamefont {P.}~\bibnamefont
  {Xu}}, \bibinfo {author} {\bibfnamefont {B.}~\bibnamefont {Li}}, \bibinfo
  {author} {\bibfnamefont {Z.~B.}\ \bibnamefont {Ren}}, \bibinfo {author}
  {\bibfnamefont {S.~H.}\ \bibnamefont {Liu}}, \ and\ \bibinfo {author}
  {\bibfnamefont {Z.~G.}\ \bibnamefont {Zuo}},\ }\bibfield  {title} {\enquote
  {\bibinfo {title} {Dynamics of a laser-induced buoyant bubble near a vertical
  rigid boundary},}\ }\href {\doibase 10.1103/PhysRevFluids.8.083601}
  {\bibfield  {journal} {\bibinfo  {journal} {Physical Review Fluids}\ }\textbf
  {\bibinfo {volume} {8}},\ \bibinfo {pages} {083601} (\bibinfo {year}
  {2023})}\BibitemShut {NoStop}%
\bibitem [{\citenamefont {Brujan}\ \emph {et~al.}(2001)\citenamefont {Brujan},
  \citenamefont {Nahen}, \citenamefont {Schmidt},\ and\ \citenamefont
  {Vogel}}]{Brujan2001}%
  \BibitemOpen
  \bibfield  {author} {\bibinfo {author} {\bibfnamefont {E.-A.}\ \bibnamefont
  {Brujan}}, \bibinfo {author} {\bibfnamefont {K.}~\bibnamefont {Nahen}},
  \bibinfo {author} {\bibfnamefont {P.}~\bibnamefont {Schmidt}}, \ and\
  \bibinfo {author} {\bibfnamefont {A.}~\bibnamefont {Vogel}},\ }\bibfield
  {title} {\enquote {\bibinfo {title} {Dynamics of laser-induced cavitation
  bubbles near elastic boundaries: influence of the elastic modulus},}\ }\href
  {\doibase 10.1017/s0022112000003335} {\bibfield  {journal} {\bibinfo
  {journal} {Journal of Fluid Mechanics}\ }\textbf {\bibinfo {volume} {433}},\
  \bibinfo {pages} {283--314} (\bibinfo {year} {2001})}\BibitemShut {NoStop}%
\bibitem [{\citenamefont {Reese}, \citenamefont {Ohl},\ and\ \citenamefont
  {Ohl}(2023)}]{Reese2023}%
  \BibitemOpen
  \bibfield  {author} {\bibinfo {author} {\bibfnamefont {H.}~\bibnamefont
  {Reese}}, \bibinfo {author} {\bibfnamefont {S.~W.}\ \bibnamefont {Ohl}}, \
  and\ \bibinfo {author} {\bibfnamefont {C.~D.}\ \bibnamefont {Ohl}},\
  }\bibfield  {title} {\enquote {\bibinfo {title} {Cavitation bubble induced
  wall shear stress on an elastic boundary},}\ }\href {\doibase
  10.1063/5.0156507} {\bibfield  {journal} {\bibinfo  {journal} {Physics of
  Fluids}\ }\textbf {\bibinfo {volume} {35}},\ \bibinfo {pages} {076122}
  (\bibinfo {year} {2023})}\BibitemShut {NoStop}%
\bibitem [{\citenamefont {Li}\ \emph {et~al.}(2019{\natexlab{a}})\citenamefont
  {Li}, \citenamefont {Zhang}, \citenamefont {Wang}, \citenamefont {Li},\ and\
  \citenamefont {Liu}}]{LiT2019Interaction}%
  \BibitemOpen
  \bibfield  {author} {\bibinfo {author} {\bibfnamefont {T.}~\bibnamefont
  {Li}}, \bibinfo {author} {\bibfnamefont {A.~M.}\ \bibnamefont {Zhang}},
  \bibinfo {author} {\bibfnamefont {S.~P.}\ \bibnamefont {Wang}}, \bibinfo
  {author} {\bibfnamefont {S.}~\bibnamefont {Li}}, \ and\ \bibinfo {author}
  {\bibfnamefont {W.~T.}\ \bibnamefont {Liu}},\ }\bibfield  {title} {\enquote
  {\bibinfo {title} {Bubble interactions and bursting behaviors near a free
  surface},}\ }\href {\doibase 10.1063/1.5088528} {\bibfield  {journal}
  {\bibinfo  {journal} {Physics of Fluids}\ }\textbf {\bibinfo {volume} {31}},\
  \bibinfo {pages} {042104} (\bibinfo {year} {2019}{\natexlab{a}})}\BibitemShut
  {NoStop}%
\bibitem [{\citenamefont {Zhang}\ \emph {et~al.}(2018)\citenamefont {Zhang},
  \citenamefont {Wang}, \citenamefont {Zhang},\ and\ \citenamefont
  {Cui}}]{Zhang2018}%
  \BibitemOpen
  \bibfield  {author} {\bibinfo {author} {\bibfnamefont {S.}~\bibnamefont
  {Zhang}}, \bibinfo {author} {\bibfnamefont {S.~P.}\ \bibnamefont {Wang}},
  \bibinfo {author} {\bibfnamefont {A.~M.}\ \bibnamefont {Zhang}}, \ and\
  \bibinfo {author} {\bibfnamefont {P.}~\bibnamefont {Cui}},\ }\bibfield
  {title} {\enquote {\bibinfo {title} {Numerical study on motion of the air-gun
  bubble based on boundary integral method},}\ }\href {\doibase
  10.1016/j.oceaneng.2018.02.008} {\bibfield  {journal} {\bibinfo  {journal}
  {Ocean Engineering}\ }\textbf {\bibinfo {volume} {154}},\ \bibinfo {pages}
  {70--80} (\bibinfo {year} {2018})}\BibitemShut {NoStop}%
\bibitem [{\citenamefont {Cui}, \citenamefont {Zhang},\ and\ \citenamefont
  {Wang}(2016)}]{CuiP2016}%
  \BibitemOpen
  \bibfield  {author} {\bibinfo {author} {\bibfnamefont {P.}~\bibnamefont
  {Cui}}, \bibinfo {author} {\bibfnamefont {A.}~\bibnamefont {Zhang}}, \ and\
  \bibinfo {author} {\bibfnamefont {S.}~\bibnamefont {Wang}},\ }\bibfield
  {title} {\enquote {\bibinfo {title} {Small-charge underwater explosion bubble
  experiments under various boundary conditions},}\ }\href {\doibase
  10.1063/1.4967700} {\bibfield  {journal} {\bibinfo  {journal} {Physics of
  Fluids}\ }\textbf {\bibinfo {volume} {28}},\ \bibinfo {pages} {117103}
  (\bibinfo {year} {2016})}\BibitemShut {NoStop}%
\bibitem [{\citenamefont {Li}\ \emph {et~al.}(2023{\natexlab{b}})\citenamefont
  {Li}, \citenamefont {Zhang}, \citenamefont {Cui}, \citenamefont {Li},\ and\
  \citenamefont {Liu}}]{LiS2023Vertical}%
  \BibitemOpen
  \bibfield  {author} {\bibinfo {author} {\bibfnamefont {S.~M.}\ \bibnamefont
  {Li}}, \bibinfo {author} {\bibfnamefont {A.~M.}\ \bibnamefont {Zhang}},
  \bibinfo {author} {\bibfnamefont {P.}~\bibnamefont {Cui}}, \bibinfo {author}
  {\bibfnamefont {S.}~\bibnamefont {Li}}, \ and\ \bibinfo {author}
  {\bibfnamefont {Y.~L.}\ \bibnamefont {Liu}},\ }\bibfield  {title} {\enquote
  {\bibinfo {title} {Vertically neutral collapse of a pulsating bubble at the
  corner of a free surface and a rigid wall},}\ }\href {\doibase
  10.1017/jfm.2023.292} {\bibfield  {journal} {\bibinfo  {journal} {Journal of
  Fluid Mechanics}\ }\textbf {\bibinfo {volume} {962}},\ \bibinfo {pages} {A28}
  (\bibinfo {year} {2023}{\natexlab{b}})}\BibitemShut {NoStop}%
\bibitem [{\citenamefont {Li}\ \emph {et~al.}(2019{\natexlab{b}})\citenamefont
  {Li}, \citenamefont {Zhang}, \citenamefont {Wang},\ and\ \citenamefont
  {Zhang}}]{LiS2019Jet}%
  \BibitemOpen
  \bibfield  {author} {\bibinfo {author} {\bibfnamefont {S.~M.}\ \bibnamefont
  {Li}}, \bibinfo {author} {\bibfnamefont {A.~M.}\ \bibnamefont {Zhang}},
  \bibinfo {author} {\bibfnamefont {Q.~X.}\ \bibnamefont {Wang}}, \ and\
  \bibinfo {author} {\bibfnamefont {S.}~\bibnamefont {Zhang}},\ }\bibfield
  {title} {\enquote {\bibinfo {title} {The jet characteristics of bubbles near
  mixed boundaries},}\ }\href {\doibase 10.1063/1.5112049} {\bibfield
  {journal} {\bibinfo  {journal} {Physics of Fluids}\ }\textbf {\bibinfo
  {volume} {31}},\ \bibinfo {pages} {107105} (\bibinfo {year}
  {2019}{\natexlab{b}})}\BibitemShut {NoStop}%
\bibitem [{\citenamefont {Zhang}\ \emph {et~al.}(2015)\citenamefont {Zhang},
  \citenamefont {Cui}, \citenamefont {Cui},\ and\ \citenamefont
  {Wang}}]{ZhangAM2015}%
  \BibitemOpen
  \bibfield  {author} {\bibinfo {author} {\bibfnamefont {A.}~\bibnamefont
  {Zhang}}, \bibinfo {author} {\bibfnamefont {P.}~\bibnamefont {Cui}}, \bibinfo
  {author} {\bibfnamefont {J.}~\bibnamefont {Cui}}, \ and\ \bibinfo {author}
  {\bibfnamefont {Q.}~\bibnamefont {Wang}},\ }\bibfield  {title} {\enquote
  {\bibinfo {title} {Experimental study on bubble dynamics subject to
  buoyancy},}\ }\href {\doibase 10.1017/jfm.2015.323} {\bibfield  {journal}
  {\bibinfo  {journal} {Journal of Fluid Mechanics}\ }\textbf {\bibinfo
  {volume} {776}},\ \bibinfo {pages} {137--160} (\bibinfo {year}
  {2015})}\BibitemShut {NoStop}%
\bibitem [{\citenamefont {Cui}\ \emph {et~al.}(2020)\citenamefont {Cui},
  \citenamefont {Chen}, \citenamefont {Wang}, \citenamefont {Zhou},\ and\
  \citenamefont {Corbett}}]{Cui2020}%
  \BibitemOpen
  \bibfield  {author} {\bibinfo {author} {\bibfnamefont {J.}~\bibnamefont
  {Cui}}, \bibinfo {author} {\bibfnamefont {Z.~P.}\ \bibnamefont {Chen}},
  \bibinfo {author} {\bibfnamefont {Q.}~\bibnamefont {Wang}}, \bibinfo {author}
  {\bibfnamefont {T.~R.}\ \bibnamefont {Zhou}}, \ and\ \bibinfo {author}
  {\bibfnamefont {C.}~\bibnamefont {Corbett}},\ }\bibfield  {title} {\enquote
  {\bibinfo {title} {Experimental studies of bubble dynamics inside a
  corner},}\ }\href {\doibase 10.1016/j.ultsonch.2019.104951} {\bibfield
  {journal} {\bibinfo  {journal} {Ultrasonics Sonochemistry}\ }\textbf
  {\bibinfo {volume} {64}},\ \bibinfo {pages} {104951} (\bibinfo {year}
  {2020})}\BibitemShut {NoStop}%
\bibitem [{\citenamefont {Cui}\ \emph {et~al.}(2023)\citenamefont {Cui},
  \citenamefont {Han}, \citenamefont {Pei},\ and\ \citenamefont
  {Wang}}]{CuiR2023}%
  \BibitemOpen
  \bibfield  {author} {\bibinfo {author} {\bibfnamefont {R.~N.}\ \bibnamefont
  {Cui}}, \bibinfo {author} {\bibfnamefont {R.}~\bibnamefont {Han}}, \bibinfo
  {author} {\bibfnamefont {S.~C.}\ \bibnamefont {Pei}}, \ and\ \bibinfo
  {author} {\bibfnamefont {S.~P.}\ \bibnamefont {Wang}},\ }\bibfield  {title}
  {\enquote {\bibinfo {title} {Jet characteristics of the three-dimensional
  explosion bubble in a compressible fluid},}\ }\href {\doibase
  10.1063/5.0163793} {\bibfield  {journal} {\bibinfo  {journal} {Physics of
  Fluids}\ }\textbf {\bibinfo {volume} {35}},\ \bibinfo {pages} {082123}
  (\bibinfo {year} {2023})}\BibitemShut {NoStop}%
\bibitem [{\citenamefont {Gonzalez-Avila}\ \emph {et~al.}(2020)\citenamefont
  {Gonzalez-Avila}, \citenamefont {van Blokland}, \citenamefont {Zeng},\ and\
  \citenamefont {Ohl}}]{Gonzalez2020}%
  \BibitemOpen
  \bibfield  {author} {\bibinfo {author} {\bibfnamefont {S.~R.}\ \bibnamefont
  {Gonzalez-Avila}}, \bibinfo {author} {\bibfnamefont {A.~C.}\ \bibnamefont
  {van Blokland}}, \bibinfo {author} {\bibfnamefont {Q.~Y.}\ \bibnamefont
  {Zeng}}, \ and\ \bibinfo {author} {\bibfnamefont {C.~D.}\ \bibnamefont
  {Ohl}},\ }\bibfield  {title} {\enquote {\bibinfo {title} {Jetting and shear
  stress enhancement from cavitation bubbles collapsing in a narrow gap},}\
  }\href {\doibase 10.1017/jfm.2019.938} {\bibfield  {journal} {\bibinfo
  {journal} {Journal of Fluid Mechanics}\ }\textbf {\bibinfo {volume} {884}},\
  \bibinfo {pages} {A23} (\bibinfo {year} {2020})}\BibitemShut {NoStop}%
\bibitem [{\citenamefont {Zhang}\ \emph {et~al.}(2021)\citenamefont {Zhang},
  \citenamefont {Lu}, \citenamefont {Zhang}, \citenamefont {Gu}, \citenamefont
  {Luo}, \citenamefont {Tong},\ and\ \citenamefont {Ren}}]{Zhang2021}%
  \BibitemOpen
  \bibfield  {author} {\bibinfo {author} {\bibfnamefont {H.~F.}\ \bibnamefont
  {Zhang}}, \bibinfo {author} {\bibfnamefont {Z.~B.}\ \bibnamefont {Lu}},
  \bibinfo {author} {\bibfnamefont {P.~H.}\ \bibnamefont {Zhang}}, \bibinfo
  {author} {\bibfnamefont {J.~Y.}\ \bibnamefont {Gu}}, \bibinfo {author}
  {\bibfnamefont {C.~H.}\ \bibnamefont {Luo}}, \bibinfo {author} {\bibfnamefont
  {Y.~Q.}\ \bibnamefont {Tong}}, \ and\ \bibinfo {author} {\bibfnamefont
  {X.~D.}\ \bibnamefont {Ren}},\ }\bibfield  {title} {\enquote {\bibinfo
  {title} {Experimental and numerical investigation of bubble oscillation and
  jet impact near a solid boundary},}\ }\href {\doibase
  10.1016/j.optlastec.2020.106606} {\bibfield  {journal} {\bibinfo  {journal}
  {Optics and Laser Technology}\ }\textbf {\bibinfo {volume} {138}},\ \bibinfo
  {pages} {106606} (\bibinfo {year} {2021})}\BibitemShut {NoStop}%
\bibitem [{\citenamefont {Zhou}\ \emph {et~al.}(2022)\citenamefont {Zhou},
  \citenamefont {Cheng}, \citenamefont {Peng}, \citenamefont {Zhang},\ and\
  \citenamefont {Shao}}]{Zhou2022}%
  \BibitemOpen
  \bibfield  {author} {\bibinfo {author} {\bibfnamefont {Z.}~\bibnamefont
  {Zhou}}, \bibinfo {author} {\bibfnamefont {X.}~\bibnamefont {Cheng}},
  \bibinfo {author} {\bibfnamefont {K.}~\bibnamefont {Peng}}, \bibinfo {author}
  {\bibfnamefont {L.}~\bibnamefont {Zhang}}, \ and\ \bibinfo {author}
  {\bibfnamefont {X.}~\bibnamefont {Shao}},\ }\bibfield  {title} {\enquote
  {\bibinfo {title} {Numerical study on the deformation of a gas bubble in
  uniform flow},}\ }\href {\doibase 10.1016/j.oceaneng.2022.111164} {\bibfield
  {journal} {\bibinfo  {journal} {Ocean Engineering}\ }\textbf {\bibinfo
  {volume} {254}},\ \bibinfo {pages} {111164} (\bibinfo {year}
  {2022})}\BibitemShut {NoStop}%
\bibitem [{\citenamefont {Mnich}\ \emph {et~al.}(2024)\citenamefont {Mnich},
  \citenamefont {Reuter}, \citenamefont {Denner},\ and\ \citenamefont
  {Ohl}}]{Mnich2024}%
  \BibitemOpen
  \bibfield  {author} {\bibinfo {author} {\bibfnamefont {D.}~\bibnamefont
  {Mnich}}, \bibinfo {author} {\bibfnamefont {F.}~\bibnamefont {Reuter}},
  \bibinfo {author} {\bibfnamefont {F.}~\bibnamefont {Denner}}, \ and\ \bibinfo
  {author} {\bibfnamefont {C.-D.}\ \bibnamefont {Ohl}},\ }\bibfield  {title}
  {\enquote {\bibinfo {title} {Single cavitation bubble dynamics in a
  stagnation flow},}\ }\href {\doibase 10.1017/jfm.2023.1048} {\bibfield
  {journal} {\bibinfo  {journal} {Journal of Fluid Mechanics}\ }\textbf
  {\bibinfo {volume} {979}},\ \bibinfo {pages} {A18} (\bibinfo {year}
  {2024})}\BibitemShut {NoStop}%
\bibitem [{\citenamefont {Su}\ \emph {et~al.}(2024)\citenamefont {Su},
  \citenamefont {Sun}, \citenamefont {Zhang}, \citenamefont {Cai},
  \citenamefont {Sun}, \citenamefont {Chen},\ and\ \citenamefont
  {Yu}}]{Su2024}%
  \BibitemOpen
  \bibfield  {author} {\bibinfo {author} {\bibfnamefont {Z.-Y.}\ \bibnamefont
  {Su}}, \bibinfo {author} {\bibfnamefont {J.}~\bibnamefont {Sun}}, \bibinfo
  {author} {\bibfnamefont {J.-W.}\ \bibnamefont {Zhang}}, \bibinfo {author}
  {\bibfnamefont {R.-Z.}\ \bibnamefont {Cai}}, \bibinfo {author} {\bibfnamefont
  {K.-F.}\ \bibnamefont {Sun}}, \bibinfo {author} {\bibfnamefont {W.-Y.}\
  \bibnamefont {Chen}}, \ and\ \bibinfo {author} {\bibfnamefont {C.-X.}\
  \bibnamefont {Yu}},\ }\bibfield  {title} {\enquote {\bibinfo {title}
  {Experimental investigation on dynamic characteristics of single bubble near
  wall in shear flow},}\ }\href {\doibase 10.1063/5.0191464} {\bibfield
  {journal} {\bibinfo  {journal} {Physics of Fluids}\ }\textbf {\bibinfo
  {volume} {36}},\ \bibinfo {pages} {033323} (\bibinfo {year}
  {2024})}\BibitemShut {NoStop}%
\bibitem [{\citenamefont {Aghdam}\ \emph {et~al.}(2015)\citenamefont {Aghdam},
  \citenamefont {Khoo}, \citenamefont {Farhangmehr},\ and\ \citenamefont
  {Shervani-Tabar}}]{Aghdam2015}%
  \BibitemOpen
  \bibfield  {author} {\bibinfo {author} {\bibfnamefont {A.~H.}\ \bibnamefont
  {Aghdam}}, \bibinfo {author} {\bibfnamefont {B.~C.}\ \bibnamefont {Khoo}},
  \bibinfo {author} {\bibfnamefont {V.}~\bibnamefont {Farhangmehr}}, \ and\
  \bibinfo {author} {\bibfnamefont {M.~T.}\ \bibnamefont {Shervani-Tabar}},\
  }\bibfield  {title} {\enquote {\bibinfo {title} {Experimental study on the
  dynamics of an oscillating bubble in a vertical rigid tube},}\ }\href
  {\doibase 10.1016/j.expthermflusci.2014.09.017} {\bibfield  {journal}
  {\bibinfo  {journal} {Experimental Thermal and Fluid Science}\ }\textbf
  {\bibinfo {volume} {60}},\ \bibinfo {pages} {299--307} (\bibinfo {year}
  {2015})}\BibitemShut {NoStop}%
\bibitem [{\citenamefont {Ni}\ \emph {et~al.}(2012)\citenamefont {Ni},
  \citenamefont {Zhang}, \citenamefont {Wang},\ and\ \citenamefont
  {Wang}}]{Ni2012}%
  \BibitemOpen
  \bibfield  {author} {\bibinfo {author} {\bibfnamefont {B.-Y.}\ \bibnamefont
  {Ni}}, \bibinfo {author} {\bibfnamefont {A.-M.}\ \bibnamefont {Zhang}},
  \bibinfo {author} {\bibfnamefont {Q.-X.}\ \bibnamefont {Wang}}, \ and\
  \bibinfo {author} {\bibfnamefont {B.}~\bibnamefont {Wang}},\ }\bibfield
  {title} {\enquote {\bibinfo {title} {Experimental and numerical study on the
  growth and collapse of a bubble in a narrow tube},}\ }\href {\doibase
  10.1007/s10409-012-0147-y} {\bibfield  {journal} {\bibinfo  {journal} {Acta
  Mechanica Sinica}\ }\textbf {\bibinfo {volume} {28}},\ \bibinfo {pages}
  {1248--1260} (\bibinfo {year} {2012})}\BibitemShut {NoStop}%
\bibitem [{\citenamefont {Wang}\ \emph {et~al.}(2024)\citenamefont {Wang},
  \citenamefont {Xu}, \citenamefont {Wang},\ and\ \citenamefont
  {Che}}]{Wang2024WallConf}%
  \BibitemOpen
  \bibfield  {author} {\bibinfo {author} {\bibfnamefont {N.}~\bibnamefont
  {Wang}}, \bibinfo {author} {\bibfnamefont {H.}~\bibnamefont {Xu}}, \bibinfo
  {author} {\bibfnamefont {T.}~\bibnamefont {Wang}}, \ and\ \bibinfo {author}
  {\bibfnamefont {Z.}~\bibnamefont {Che}},\ }\bibfield  {title} {\enquote
  {\bibinfo {title} {Wall confinement effects on the dynamics of cavitation
  bubbles in thin tubes},}\ }\href {\doibase 10.1063/5.0196787} {\bibfield
  {journal} {\bibinfo  {journal} {Physics of Fluids}\ }\textbf {\bibinfo
  {volume} {36}},\ \bibinfo {pages} {042101} (\bibinfo {year}
  {2024})}\BibitemShut {NoStop}%
\bibitem [{\citenamefont {Li}\ \emph {et~al.}(2024)\citenamefont {Li},
  \citenamefont {Zhou}, \citenamefont {Luo}, \citenamefont {Xu}, \citenamefont
  {Zhai}, \citenamefont {Qu},\ and\ \citenamefont {Zou}}]{LiJie2024Collapsing}%
  \BibitemOpen
  \bibfield  {author} {\bibinfo {author} {\bibfnamefont {J.}~\bibnamefont
  {Li}}, \bibinfo {author} {\bibfnamefont {M.}~\bibnamefont {Zhou}}, \bibinfo
  {author} {\bibfnamefont {J.}~\bibnamefont {Luo}}, \bibinfo {author}
  {\bibfnamefont {W.}~\bibnamefont {Xu}}, \bibinfo {author} {\bibfnamefont
  {Y.}~\bibnamefont {Zhai}}, \bibinfo {author} {\bibfnamefont {T.}~\bibnamefont
  {Qu}}, \ and\ \bibinfo {author} {\bibfnamefont {L.}~\bibnamefont {Zou}},\
  }\bibfield  {title} {\enquote {\bibinfo {title} {Collapsing behavior of
  spark-induced cavitation bubble in rigid tube},}\ }\href {\doibase
  10.1016/j.ultsonch.2024.106791} {\bibfield  {journal} {\bibinfo  {journal}
  {Ultrasonics Sonochemistry}\ ,\ \bibinfo {pages} {106791}} (\bibinfo {year}
  {2024})}\BibitemShut {NoStop}%
\bibitem [{\citenamefont {Ory}\ \emph {et~al.}(2000)\citenamefont {Ory},
  \citenamefont {Yuan}, \citenamefont {Prosperetti}, \citenamefont {Popinet},\
  and\ \citenamefont {Zaleski}}]{Ory2000}%
  \BibitemOpen
  \bibfield  {author} {\bibinfo {author} {\bibfnamefont {E.}~\bibnamefont
  {Ory}}, \bibinfo {author} {\bibfnamefont {H.}~\bibnamefont {Yuan}}, \bibinfo
  {author} {\bibfnamefont {A.}~\bibnamefont {Prosperetti}}, \bibinfo {author}
  {\bibfnamefont {S.}~\bibnamefont {Popinet}}, \ and\ \bibinfo {author}
  {\bibfnamefont {S.}~\bibnamefont {Zaleski}},\ }\bibfield  {title} {\enquote
  {\bibinfo {title} {Growth and collapse of a vapor bubble in a narrow tube},}\
  }\href {\doibase 10.1063/1.870381} {\bibfield  {journal} {\bibinfo  {journal}
  {Physics of Fluids}\ }\textbf {\bibinfo {volume} {12}},\ \bibinfo {pages}
  {1268--1277} (\bibinfo {year} {2000})}\BibitemShut {NoStop}%
\bibitem [{\citenamefont {Yuan}\ and\ \citenamefont
  {Prosperetti}(1999)}]{Yuan1999}%
  \BibitemOpen
  \bibfield  {author} {\bibinfo {author} {\bibfnamefont {H.}~\bibnamefont
  {Yuan}}\ and\ \bibinfo {author} {\bibfnamefont {A.}~\bibnamefont
  {Prosperetti}},\ }\bibfield  {title} {\enquote {\bibinfo {title} {The pumping
  effect of growing and collapsing bubbles in a tube},}\ }\href {\doibase
  10.1088/0960-1317/9/4/318} {\bibfield  {journal} {\bibinfo  {journal}
  {Journal of Micromechanics and Microengineering}\ }\textbf {\bibinfo {volume}
  {9}},\ \bibinfo {pages} {402--413} (\bibinfo {year} {1999})}\BibitemShut
  {NoStop}%
\bibitem [{\citenamefont {Wang}\ \emph {et~al.}(2019)\citenamefont {Wang},
  \citenamefont {Wang}, \citenamefont {Zhang},\ and\ \citenamefont
  {Stride}}]{Wang2019Exp}%
  \BibitemOpen
  \bibfield  {author} {\bibinfo {author} {\bibfnamefont {S.~P.}\ \bibnamefont
  {Wang}}, \bibinfo {author} {\bibfnamefont {Q.~X.}\ \bibnamefont {Wang}},
  \bibinfo {author} {\bibfnamefont {A.~M.}\ \bibnamefont {Zhang}}, \ and\
  \bibinfo {author} {\bibfnamefont {E.}~\bibnamefont {Stride}},\ }\bibfield
  {title} {\enquote {\bibinfo {title} {Experimental observations of the
  behaviour of a bubble inside a circular rigid tube},}\ }\href {\doibase
  10.1016/j.ijmultiphaseflow.2019.103096} {\bibfield  {journal} {\bibinfo
  {journal} {International Journal of Multiphase Flow}\ }\textbf {\bibinfo
  {volume} {121}},\ \bibinfo {pages} {103096} (\bibinfo {year}
  {2019})}\BibitemShut {NoStop}%
\bibitem [{\citenamefont {Ren}\ \emph {et~al.}(2022)\citenamefont {Ren},
  \citenamefont {Li}, \citenamefont {Xu}, \citenamefont {Wakata}, \citenamefont
  {Liu}, \citenamefont {Sun}, \citenamefont {Zuo},\ and\ \citenamefont
  {Liu}}]{ren22}%
  \BibitemOpen
  \bibfield  {author} {\bibinfo {author} {\bibfnamefont {Z.~B.}\ \bibnamefont
  {Ren}}, \bibinfo {author} {\bibfnamefont {B.}~\bibnamefont {Li}}, \bibinfo
  {author} {\bibfnamefont {P.}~\bibnamefont {Xu}}, \bibinfo {author}
  {\bibfnamefont {Y.}~\bibnamefont {Wakata}}, \bibinfo {author} {\bibfnamefont
  {J.}~\bibnamefont {Liu}}, \bibinfo {author} {\bibfnamefont {C.}~\bibnamefont
  {Sun}}, \bibinfo {author} {\bibfnamefont {Z.~G.}\ \bibnamefont {Zuo}}, \ and\
  \bibinfo {author} {\bibfnamefont {S.~H.}\ \bibnamefont {Liu}},\ }\bibfield
  {title} {\enquote {\bibinfo {title} {Cavitation bubble dynamics in a
  funnel-shaped tube},}\ }\href {\doibase 10.1063/5.0107436} {\bibfield
  {journal} {\bibinfo  {journal} {Physics of Fluids}\ }\textbf {\bibinfo
  {volume} {34}},\ \bibinfo {pages} {093313} (\bibinfo {year}
  {2022})}\BibitemShut {NoStop}%
\bibitem [{\citenamefont {Nagargoje}\ and\ \citenamefont
  {Gupta}(2023)}]{nagargoje23}%
  \BibitemOpen
  \bibfield  {author} {\bibinfo {author} {\bibfnamefont {M.~S.}\ \bibnamefont
  {Nagargoje}}\ and\ \bibinfo {author} {\bibfnamefont {R.}~\bibnamefont
  {Gupta}},\ }\bibfield  {title} {\enquote {\bibinfo {title} {Experimental
  investigations on the bubble dynamics in a symmetric bifurcating channel},}\
  }\href {\doibase 10.1016/j.ijmultiphaseflow.2022.104318} {\bibfield
  {journal} {\bibinfo  {journal} {International Journal of Multiphase Flow}\
  }\textbf {\bibinfo {volume} {159}},\ \bibinfo {pages} {104318} (\bibinfo
  {year} {2023})}\BibitemShut {NoStop}%
\bibitem [{\citenamefont {Luo}\ \emph {et~al.}(2022)\citenamefont {Luo},
  \citenamefont {Chen}, \citenamefont {Xiao}, \citenamefont {Yao},\ and\
  \citenamefont {Liu}}]{Luo2022}%
  \BibitemOpen
  \bibfield  {author} {\bibinfo {author} {\bibfnamefont {X.}~\bibnamefont
  {Luo}}, \bibinfo {author} {\bibfnamefont {T.}~\bibnamefont {Chen}}, \bibinfo
  {author} {\bibfnamefont {W.}~\bibnamefont {Xiao}}, \bibinfo {author}
  {\bibfnamefont {X.}~\bibnamefont {Yao}}, \ and\ \bibinfo {author}
  {\bibfnamefont {J.}~\bibnamefont {Liu}},\ }\bibfield  {title} {\enquote
  {\bibinfo {title} {The dynamics of a bubble in the internal fluid flow of a
  pipeline},}\ }\href {\doibase 10.1063/5.0112496} {\bibfield  {journal}
  {\bibinfo  {journal} {Physics of Fluids}\ }\textbf {\bibinfo {volume} {34}},\
  \bibinfo {pages} {117103} (\bibinfo {year} {2022})}\BibitemShut {NoStop}%
\bibitem [{\citenamefont {Feng}(2009)}]{Feng2009}%
  \BibitemOpen
  \bibfield  {author} {\bibinfo {author} {\bibfnamefont {J.~Q.}\ \bibnamefont
  {Feng}},\ }\bibfield  {title} {\enquote {\bibinfo {title} {A long gas bubble
  moving in a tube with flowing liquid},}\ }\href {\doibase
  10.1016/j.ijmultiphaseflow.2009.03.012} {\bibfield  {journal} {\bibinfo
  {journal} {International journal of multiphase flow}\ }\textbf {\bibinfo
  {volume} {35}},\ \bibinfo {pages} {738--746} (\bibinfo {year}
  {2009})}\BibitemShut {NoStop}%
\bibitem [{\citenamefont {Feng}(2010)}]{Feng2010}%
  \BibitemOpen
  \bibfield  {author} {\bibinfo {author} {\bibfnamefont {J.~Q.}\ \bibnamefont
  {Feng}},\ }\bibfield  {title} {\enquote {\bibinfo {title} {Steady
  axisymmetric motion of a small bubble in a tube with flowing liquid},}\
  }\href {\doibase 10.1098/rspa.2009.0288} {\bibfield  {journal} {\bibinfo
  {journal} {Proceedings of the Royal Society A: Mathematical, Physical and
  Engineering Sciences}\ }\textbf {\bibinfo {volume} {466}},\ \bibinfo {pages}
  {549--562} (\bibinfo {year} {2010})}\BibitemShut {NoStop}%
\bibitem [{\citenamefont {Weller}\ \emph {et~al.}(1998)\citenamefont {Weller},
  \citenamefont {Tabor}, \citenamefont {Jasak},\ and\ \citenamefont
  {Fureby}}]{Weller1998}%
  \BibitemOpen
  \bibfield  {author} {\bibinfo {author} {\bibfnamefont {H.~G.}\ \bibnamefont
  {Weller}}, \bibinfo {author} {\bibfnamefont {G.}~\bibnamefont {Tabor}},
  \bibinfo {author} {\bibfnamefont {H.}~\bibnamefont {Jasak}}, \ and\ \bibinfo
  {author} {\bibfnamefont {C.}~\bibnamefont {Fureby}},\ }\bibfield  {title}
  {\enquote {\bibinfo {title} {A tensorial approach to computational continuum
  mechanics using object-oriented techniques},}\ }\href {\doibase
  10.1063/1.168744} {\bibfield  {journal} {\bibinfo  {journal} {Computers in
  Physics}\ }\textbf {\bibinfo {volume} {12}},\ \bibinfo {pages} {620--631}
  (\bibinfo {year} {1998})}\BibitemShut {NoStop}%
\bibitem [{\citenamefont {Brackbill}, \citenamefont {Kothe},\ and\
  \citenamefont {Zemach}(1992)}]{Brackbill1992}%
  \BibitemOpen
  \bibfield  {author} {\bibinfo {author} {\bibfnamefont {J.~U.}\ \bibnamefont
  {Brackbill}}, \bibinfo {author} {\bibfnamefont {D.~B.}\ \bibnamefont
  {Kothe}}, \ and\ \bibinfo {author} {\bibfnamefont {C.}~\bibnamefont
  {Zemach}},\ }\bibfield  {title} {\enquote {\bibinfo {title} {A continuum
  method for modeling surface tension},}\ }\href {\doibase
  10.1016/0021-9991(92)90240-Y} {\bibfield  {journal} {\bibinfo  {journal}
  {Journal of Computational Physics}\ }\textbf {\bibinfo {volume} {100}},\
  \bibinfo {pages} {335--354} (\bibinfo {year} {1992})}\BibitemShut {NoStop}%
\bibitem [{\citenamefont {Koch}\ \emph {et~al.}(2016)\citenamefont {Koch},
  \citenamefont {Lechner}, \citenamefont {Reuter}, \citenamefont {Kohler},
  \citenamefont {Mettin},\ and\ \citenamefont {Lauterborn}}]{Koch2016}%
  \BibitemOpen
  \bibfield  {author} {\bibinfo {author} {\bibfnamefont {M.}~\bibnamefont
  {Koch}}, \bibinfo {author} {\bibfnamefont {C.}~\bibnamefont {Lechner}},
  \bibinfo {author} {\bibfnamefont {F.}~\bibnamefont {Reuter}}, \bibinfo
  {author} {\bibfnamefont {K.}~\bibnamefont {Kohler}}, \bibinfo {author}
  {\bibfnamefont {R.}~\bibnamefont {Mettin}}, \ and\ \bibinfo {author}
  {\bibfnamefont {W.}~\bibnamefont {Lauterborn}},\ }\bibfield  {title}
  {\enquote {\bibinfo {title} {Numerical modeling of laser generated cavitation
  bubbles with the finite volume and volume of fluid method, using
  {OpenFOAM}},}\ }\href {\doibase 10.1016/j.compfluid.2015.11.008} {\bibfield
  {journal} {\bibinfo  {journal} {Computers \& Fluids}\ }\textbf {\bibinfo
  {volume} {126}},\ \bibinfo {pages} {71--90} (\bibinfo {year}
  {2016})}\BibitemShut {NoStop}%
\end{thebibliography}%
\end{document}